 \documentstyle[onecolumn,epsf]{mn}

\begin{document}
\title[On the contribution of density perturbations and gravitational waves...]
{On the contribution of density perturbations and gravitational waves
to the lower order multipoles of the Cosmic Microwave Background Radiation}
\author[A. Dimitropoulos \& L.P.Grishchuk]{A. Dimitropoulos$^{\footnotesize{(1)}}$\thanks{e-mail address:
A.Dimitropoulos@astro.cf.ac.uk} and 
L.P.Grishchuk$^{\footnotesize(1,2)}$
\thanks{e-mail address: Leonid.Grishchuk@astro.cf.ac.uk}\\
$(1)$Cardiff University, Department of Physics and Astronomy, P.O.Box 913, 
Cardiff, CF24 3YB, United Kingdom\\
$(2)$Sternberg Astronomical Institute, Moscow University, Moscow 119899, 
Russia}
\maketitle

\begin{abstract}
The important studies of Peebles, and Bond and Efstathiou have led to
the formula $C_{\ell} = const/[ \ell(\ell +1) ]$ aimed at describing the 
lower order multipoles of the CMBR temperature variations caused by 
density perturbations with the flat spectrum. 
Clearly, this formula requires amendments, as it predicts an infinitely large 
monopole $C_{0}$, and a dipole moment $C_{1}$ only $6/2$ times larger than
the quadrupole $C_{2}$, both predictions in conflict with observations. We 
restore the terms 
omitted in the course of the derivation of this formula, and arrive at a new 
expression. According to the corrected formula, the monopole moment is finite 
and small, while the dipole moment is sensitive to short-wavelength 
perturbations, 
and numerically much larger than the quadrupole, as one would expect on 
physical grounds. 
At the same time, the function $\ell (\ell +1)C_{\ell}$ deviates 
from a horizontal line and grows with $\ell$, for $\ell \geq 2$. We show 
that the inclusion of the modulating (transfer) function terminates the growth 
and forms the first peak, recently observed. We fit the theoretical curves to 
the 
position and height of the first peak, as well as to the observed dipole, 
varying 
three parameters: red-shift at decoupling, red-shift at matter-radiation
equality, and slope of the primordial spectrum. It appears that there
is always a deficit, as compared with the COBE observations, at small 
multipoles, $\ell \sim 10$. We demonstrate that a reasonable and theoretically 
expected amount of gravitational waves bridges this gap at small multipoles, 
leaving the other fits as good as before. We show that the observationally 
acceptable 
models permit somewhat `blue' primordial spectra. This allows one to avoid
the infra-red divergence of cosmological perturbations, which is otherwise 
present.     
\end{abstract}


\section{Introduction}

The important studies of Peebles \cite{Peeb,Pebook}
and Bond and 
Efstathiou \cite{BondEf,Efst} have led to 
the formula
for the multipole moments of the CMB radiation, $C_{\ell}$, caused by small
density perturbations with flat spectrum
(also known as Harrison--Zel'dovich--Peebles, or scale-invariant 
spectrum; spectral index ${\mathrm{n}}=1$):
\begin{equation}
C_{\ell} = \frac{const}{\ell (\ell +1)}. \label{eq:one}
\end{equation}
This formula aims to describe the lower order multipoles of the CMBR 
variations in a 
spatially flat universe. The general acceptance of this result is reflected 
in the 
fact that the $C_{\ell}$ distributions are usually plotted in terms of the 
function 
$\ell(\ell +1)C_{\ell}$ versus $\ell$. In these units, formula (\ref{eq:one}) 
is conveniently
plotted as a horizontal line. 

Clearly, formula (\ref{eq:one}) was derived under some assumptions, and this 
derivation
requires amendments. Indeed, let us start from the monopole moment $C_{0}$,
setting $\ell =0$ in this formula. Expression (\ref{eq:one}) predicts 
infinity for $C_{0}$, 
in conflict with observations. Surely, it would be difficult to 
observationally distinguish 
a monopole component caused by perturbations from the isotropic unperturbed 
temperature 
$T_{0}$, but it is strange that a theoretical formula gives infinity for an 
effect produced 
by small perturbations. Let us now consider the dipole, $\ell =1$, and the
quadrupole, $\ell=2$, moments. According to (\ref{eq:one}), the quantities 
$C_{1}$ and $C_{2}$ 
are numerically of the same order of magnitude, $C_{1} = (6/2) C_{2}$. 
However, we know 
from observations that the observed dipole variation of the temperature is 
about 100 times 
larger than the observed quadrupole variation.

The predictions of formula (\ref{eq:one}) also seem to disagree with 
independent theoretical 
studies. According to Grishchuk \& Zel'dovich \shortcite{GriZe}, in the limit 
of long-wavelength 
perturbations, which 
normally provide dominant contributions to the $\ell =0$ and $\ell= 2$ 
multipoles, 
the monopole and quadrupole variations of the CMBR temperature should be 
small and approximately 
equal. At the same time, the dipole variation should be suppressed, as 
compared with the monopole 
and quadrupole, in proportion to the ratio $l_{{\mathrm{H}}}/ \lambda$, where 
$l_{{\mathrm{H}}}$ is 
the Hubble radius 
and $\lambda$ is the perturbation's wavelength, 
$\lambda \gg l_{{\mathrm{H}}} $. The 
multipole moments 
$C_{\ell}$, being quadratic 
in the temperature variations, contain this factor in the second power. The 
factor 
$l_{{\mathrm{H}}}/\lambda$ becomes much larger than 1 in the opposite limit of 
short-wavelength perturbations, 
$\lambda \ll l_{{\mathrm{H}}}$. The dipole component is, thus, expected to be 
very 
sensitive to short waves, 
and its value should eventually exceed the values of the monopole and 
quadrupole components, as 
soon as the contribution of short-wavelength perturbations is also included. 
However, formula 
(\ref{eq:one}) suggests results differing from these conclusions. The 
discrepancy is especially 
troublesome, as the derivation of (\ref{eq:one}) seems to have taken into
account the perturbations of all 
wavelengths. Indeed, this formula arises from the table integral 
(discussed below in Section \ref{sec-THIRD})
\[ \int_{0}^{\infty}[j_{\ell}(n\xi)]^2 \frac{{\mathrm{d}}n}{n} = 
\frac{1}{2\ell (\ell +1)},\]
where the integration over wave-numbers $n$ ranges formally from zero to 
infinity. Therefore, according to this integral, 
the 
very short-wavelength 
perturbations have also been taken into account in the calculation of 
all $C_{\ell}$'s, including the dipole $C_{1}$. 

The demonstrated deficiencies of formula (\ref{eq:one}) for $\ell =0$ and 
$\ell =1$ indicate
that there could also be significant deviations from the values suggested by 
(\ref{eq:one}) 
in the case of $\ell=2$ and higher order multipoles. This consideration has 
prompted us to 
reanalyse the derivation of formula (\ref{eq:one}). In Section 
\ref{sec-FIRST} we introduce the necessary 
notations and remind the reader the basics of the Sachs--Wolfe result 
\cite{SaWo}. For the 
purpose of later comparison with observations, we consider primordial spectra 
with 
arbitrary spectral indices, so that the flat spectrum ${\mathrm{n}} = 1$ is 
simply a particular 
case. Section \ref{sec-SECOND} summarises the statistical assumptions and 
defines the multipole moments. In 
Section \ref{sec-THIRD} we identify the terms that were dropped out in the 
course 
of the original derivation of 
formula (\ref{eq:one}). We show the importance of these terms and restore 
them. After this 
correction, the numerical values of the monopole and quadrupole moments 
become comparable to each 
other, while the dipole moment becomes sensitive to short-wavelength 
perturbations, as it should. 
At the same time, the correct formula for $C_{\ell}$ requires the function 
$\ell (\ell+1)C_{\ell}$ 
to deviate from the horizontal line and grow with $\ell$ for $\ell \geq 2$. 
Formally, this
growth would be unlimited if one were allowed to use the primordial power-law 
spectrum for arbitrarily 
large values of the wavenumber $n$. However, it is known that the primordial 
power spectrum of density 
perturbations should experience a turnover, caused by the transition of the 
cosmological
expansion from the radiation-dominated to the matter-dominated regime 
\cite{ZelNo,Pebook}. By the 
time of decoupling, the primordial spectrum gets modified. Mathematically, 
the original power spectrum 
gets multiplied by the modulating function (transfer function). The 
modulating function is practically 
unity in the spectral region of long-wavelength perturbations, but it 
strongly suppresses the power in 
short-wavelength perturbations. As a result, the growth of the function 
$\ell (\ell+1)C_{\ell}$ 
terminates and is being followed by its decline at multipoles, roughly, 
around $\ell=200$. This causes 
the function $\ell (\ell+1) C_{\ell}$ to form the first peak, recently 
observed \cite{Hana,deBer,Maus}. 
The following (less prominent) peaks of the modulating function are 
responsible for the further peaks of 
the function $\ell (\ell +1)C_{\ell}$. In Section \ref{sec-FOURTH} we analyse 
in detail the form of the modulating 
function and derive the possible (theoretical) positions and heights of the 
first peak. We think that 
the origin of the subsequent peaks in the $\ell$-space is due to the 
modulation of the metric power spectrum of 
density perturbations (oscillations in the $n$-space). This 
modulation, in its turn, entirely 
arises due to the standing-wave character of the initial density 
perturbations. Travelling sound waves 
in the radiation-dominated era would not have produced Sakharov oscillations 
at the time of decoupling, 
as was explained long ago (see Section 11.6 in Zel'dovich \& Novikov 
\shortcite{ZelNo}). The recent paper \cite{Miller}
may be providing support to this interpretation of the peaks. However,
the origin and position of secondary peaks is not the subject of this
paper, so we leave the door open for alternative interpretations. The full 
computation of all multipoles is 
presented in Section \ref{sec-FIFTH}. We replace the existing uncertainties 
about the matter content of the Universe 
by the two phenomenological parameters: the decoupling red-shift 
$z_{{\mathrm{dec}}}$ 
and the matter-radiation equality 
red-shift $z_{{\mathrm{eq}}}$. We also vary the third parameter, $\beta$, 
which 
characterises the spectral slope of 
the primordial perturbations. (The flat spectrum ${\mathrm{n}} = 1$ 
corresponds to $\beta = -2$.) We show 
how the distributions of multipole moments depend on the choice of these 
parameters. Since we concentrate 
on anisotropies of gravitational origin, the chemical composition of matter 
in the Universe is unimportant, 
as long as the perturbations can be described by solutions to the perturbed 
Einstein equations at the
radiation-dominated stage (equation of state $p = \epsilon/3$) followed by 
the matter-dominated stage 
(equation of state $p=0$). Section \ref{sec-SIXTH} gives a preliminary 
comparison of the calculated multipole moments 
(including the dipole) with the available CMBR observations. We require the 
theoretical (statistical) dipole, 
as well as the location and the height of the first peak, to be as close as 
possible to their observed values.  
It appears, then, that the values of the COBE-observed \cite{Smoot} 
multipoles, around $\ell =10$, 
are somewhat larger than the theoretical curves for $\ell (\ell +1)C_{\ell}$ 
would require. The, so far 
neglected, intrinsic anisotropies (temperature variations at the last 
scattering surface) can only increase the 
height of the first peak making this discrepancy larger. The most natural way 
of removing this discrepancy is 
to take into account gravitational waves. They mostly contribute to the 
multipoles from $\ell=2$ 
to approximately $\ell=30$, without changing the dipole, and not seriously 
affecting the location and height of 
the peak. Section \ref{sec-SEVENTH} describes the contribution of 
gravitational waves. In effect, we calculate the amount 
of gravitational waves needed in order to raise the `plateau' at lower 
$\ell$'s to the actually observed 
level. In Section \ref{sec-EIGHTH} we again compare the theory with 
observations, this time including gravitational 
waves, and discuss the values of the parameters $z_{{\mathrm{dec}}}$, 
$z_{{\mathrm{eq}}}$, $\beta$ 
that produce distributions
$\ell (\ell +1) C_{\ell}$ in better agreement with CMBR observations. We 
argue that a slightly larger amount 
of gravitational waves allows the primordial spectrum to be somewhat `blue' 
(${\mathrm{n}} > 1$ or 
$\beta > -2$) instead of being flat. The $C_{\ell}$'s produced by such 
`blue' spectra are still consistent 
with the discussed observations. We consider this fact as a satisfactory 
conclusion, since, from the theoretical
standpoint, the flat spectrum and `red' spectra (${\mathrm{n}} < 1$ or 
$\beta < -2$) are undesirable. Indeed, 
the mean square values of the metric (gravitational field) fluctuations are 
logarithmically divergent for the 
${\mathrm{n}} = 1$ case and power-law divergent for all ${\mathrm{n}} < 1$, 
in the limit of long waves. The 
removal of these divergences would require extra assumptions. Appendix A 
contains some technical details of the 
derivation of the corrected formula for $C_{\ell}$.

\section{The Sachs-Wolfe result and growing density perturbations}
\label{sec-FIRST}

Following Landau \& Lifshitz \shortcite{LaLif}, 
Sachs \& Wolfe \shortcite{SaWo}, we describe a FLRW universe and the 
perturbed gravitational field by the line element
\[ {\mathrm{d}}s^{2} = a^{2}[{\mathrm{d}}\eta ^{2} - (\delta_{ij} + h_{ij}) 
{\mathrm{d}}x^{i}\ {\mathrm{d}}x^{j}], \]
where $h_{ij}$ refers to density perturbations or gravitational waves. In the 
matter-dominated era (equation of 
state $p=0$) the scale factor behaves as $a(\eta) \propto \eta^2$, which we 
write as
\[  a_{{\mathrm{m}}}\!(\eta) = 2l_{{\mathrm{H}}}\ 
(\eta - \eta_{{\mathrm{m}}})^{2}, \] 
where $l_{{\mathrm{H}}}$ is the Hubble radius at the present time 
$\eta = \eta_{{\mathrm{R}}}$. The 
constant $\eta_{{\mathrm{m}}}$ is needed for 
further considerations. It is convenient to choose 
$\eta_{{\mathrm{R}}} -\eta_{{\mathrm{m}}} =1$. 
With this convention, the wave-length 
equal to the Hubble radius today corresponds to the wave-number 
$n_{{\mathrm{H}}}=4\pi$. 
In the matter-dominated era, the 
employed coordinate system is both synchronous and comoving - the most 
convenient choice \cite{LaLif,SaWo}. 
The elements of plasma emitting the CMB photons, as well as the observer, 
are `frozen' in the perturbed deforming 
matter and, 
by the definition of the comoving coordinate system, have zero velocities 
with respect to the matter fluid. 

Let the direction of observation of a photon emitted at 
$\eta = \eta_{{\mathrm{E}}}$ be 
characterised by a unit vector 
   \begin{equation}
      {\mathbf{e}} = (\sin\!\theta \ \cos\!\phi ,\ \  \sin\!\theta \ 
\sin\!\phi ,\ \  \cos\!\theta) . \label{eq:unitvector}
   \end{equation}
The relative temperature variation $\delta T /T$ seen in a given direction 
$(\theta,\phi)$ is described by the 
integral \cite{SaWo} 
   \begin{equation}
      \frac{\delta T}{T}\!({\mathbf{e}}) = \frac{1}{2} \int_{0}^{\xi}
\left.\frac{\partial h_{ij}}{\partial \eta} e^{i} e^{j}\right|_{{\mathbf{x}}=
{\mathbf{e}}w} {\mathrm{d}}w , \label{eq:SWformula}
   \end{equation} 
where $\xi = \eta_{{\mathrm{R}}} - \eta_{{\mathrm{E}}}$ and the integration 
is performed along the 
path $\eta = \eta_R -w$, ${\mathbf{x}} = {\mathbf{e}} w$. 
The quantity $\xi$ is related to the red-shift $z_{{\mathrm{dec}}}$ at 
decoupling as
\begin{equation}
\xi =1 - (1+ z_{{\mathrm{dec}}})^{-1/2}. \label{eq:redxi} 
\end{equation}
Typically, the quantity $\xi$ is very close to 1. The fact that the photons 
coming from different directions might 
have been emitted at slightly different $\eta_{{\mathrm{E}}}$'s is irrelevant 
for this 
calculation, as the integrand in 
expression (\ref{eq:SWformula}) is already a small quantity of first order. A 
first order variation of the upper 
limit of integration will change the result only in second order of 
smallness. However, a small initial variation 
of the temperature over the last scattering surface will be reflected in an 
additive term to the integral 
(\ref{eq:SWformula}). This term would be effective even in the absence of 
the intervening gravitational field $h_{ij}$ 
along the photon's path, and it can be called the intrinsic anisotropy 
$(\delta T / T)_{{\mathrm{in}}}$. We discuss this term later, 
in Section \ref{sec-FIFTH}. As long as further scattering and absorption of 
CMB photons can be neglected, the integral 
(\ref{eq:SWformula}) plus the $(\theta,\phi)$-dependent initial condition 
$(\delta T / T)_{{\mathrm{in}}}$ fully describe the 
temperature variations at the point of reception.  

We now consider density perturbations. By the time of decoupling, one can 
normally neglect one of the two 
linearly-independent solutions to the perturbed equations, and work with the 
so-called growing solution. In the case of 
density perturbations at the matter-dominated stage, the growing solution is
     \begin{equation}
h_{ij}\!(\eta,{\mathbf{x}}) = B\!({\mathbf{x}}) \delta_{ij} + 
\frac{(\eta - \eta _{{\mathrm{m}}})^{2}}{10} B\!({\mathbf{x}})_{,ij} .   
\label{eq:firstpertdp}
     \end{equation}
Using formula (\ref{eq:firstpertdp}) in the integral (\ref{eq:SWformula}), 
one obtains \cite{SaWo}
  \begin{equation}
    \frac{\delta T}{T}\!({\mathbf{e}})  =  \frac{1}{10} \left. B_{,f} e^{f} 
\right|_{{\mathrm{R}}} - \frac{1}{10}(1-\xi) 
\left. B_{,f} e^{f} \right|_{{\mathrm{E}}} + 
\frac{1}{10} B|_{{\mathrm{R}}} - \frac{1}{10} B|_{{\mathrm{E}}}. 
\label{eq:tempanisfirst}
  \end{equation}
The exceptional property of the solution (\ref{eq:firstpertdp}) is that the 
integrand becomes a total derivative, so the
integral (\ref{eq:SWformula}) reduces 
to the values of the under-integral quantity at the end-points of 
integration. This property is responsible for formula 
(\ref{eq:tempanisfirst}). In the general case, this is not 
true, and is only 
approximately true for other types of perturbations when they are taken in 
the long-wavelength limit \cite{GriZe}. In view 
of the already existing disorder in the literature with respect to the naming 
and interpretation of different parts of 
the Sachs--Wolfe work, we simply call the terms in (\ref{eq:tempanisfirst}), 
in order of their 
appearance, as term I, term II, term III, and term IV. Term I is purely 
dipolar in its angular structure. Term II 
consists of all spherical harmonics, including the monopole. Term III has no 
angular dependence: it only contributes to 
the monopole. Term IV consists of all spherical harmonics, including the 
monopole.

One normally assumes that cosmological perturbations can be decomposed over 
spatial Fourier components with arbitrary 
wave-vectors ${\mathbf{n}}$. In our case, this amounts to the decomposition 
of $B({\mathbf{x}})$ as 
     \begin{equation}
         B\!({\mathbf{x}}) \equiv \frac{1}{(2\pi)^{3/2}}
\int_{-\infty}^{+\infty} {\mathrm{d}}^{3}{\mathbf{n}}\ 
\left( B_{{\mathbf{n}}}\ 
{\mathrm{e}}^{{\mathrm{i}}{\mathbf{n\cdot x}}} + B^{\ast}_{{\mathbf{n}}}\ 
{\mathrm{e}}^{-{\mathrm{i}}{\mathbf{n\cdot x}}} 
\right).     \label{eq:Bfourrier}
     \end{equation}
Since $B\!({\mathbf{x}})$ is a real function, we have included a complex 
conjugate part to make the expression manifestly 
real. Using the notation
       \[   G\!({\mathbf{e}}) \equiv \frac{1}{(2\pi)^{3/2}}
\ \int_{-\infty}^{+\infty} {\mathrm{d}}^{3}{\mathbf{n}}\ B_{{\mathbf{n}}}\ 
\left\{{\mathrm{i}} {\mathbf{n\cdot e}} \left[1 - (1 -\xi)
{\mathrm{e}}^{{\mathrm{i}} {\mathbf{n\cdot e}} \xi} \right] + 
\left(1 - {\mathrm{e}}^{{\mathrm{i}} {\mathbf{n\cdot e}} \xi} \right)  \right\}
,\]      
the temperature variation (\ref{eq:tempanisfirst}) can be written as 
      \begin{equation}
         \frac{\delta T}{T}\!({\mathbf{e}})  = \frac{1}{10} 
[ G\!({\mathbf{e}}) +  G^{\ast}\!({\mathbf{e}})]. \label{eq:tempanisG}
      \end{equation}
One can expand $G\!({\mathbf{e}})$ over complex spherical harmonics:  
      \begin{equation}
         G\!({\mathbf{e}}) = \sum_{\ell =0}^{\infty}\sum_{m=-\ell}^{\ell}  
g_{\ell m}  Y_{\ell m}\!(\theta ,\phi).  \label{eq:Gspherdef}
      \end{equation}
To do this explicitly, it is convenient to write the wave vector 
${\mathbf{n}}$ in the form
         \begin{equation}
           {\mathbf{n}} = (n \sin\!\Theta\ \cos\!\Phi ,\ \ n \sin\!\Theta\ 
\sin\!\Phi ,\ \ n \cos\!\Theta)  \label{eq:wavevector}
         \end{equation}
and use the expansion (see, for example, formula (16.127) of 
Jackson \shortcite{Jack}):
       \begin{equation}
          \exp \{ {\mathrm{i}} {\mathbf{n\cdot x}} \} = 4\pi\ 
\sum_{\ell =0}^{\infty} 
\sum_{m=-\ell}^{\ell} {\mathrm{i}}^{\ell}\ j_{\ell}\!(nx)\ 
Y_{\ell m}^{\ast}\!(\Theta ,\Phi)\ Y_{\ell m}\!(\theta ,\phi).       
\label{eq:A17}
       \end{equation}
As a result, one finds that
\begin{description}
\item[term I] produces a purely dipolar variation
\[
 {\mathrm{i}} {\mathbf{n\cdot e}} =  {\mathrm{i}}\ \frac{4\pi}{3}\ 
n\sum_{f=-1}^{1} 
Y_{1f}^{\ast}(\Theta ,\Phi)\ Y_{1f}\!(\theta ,\phi) = 
4\pi\ \sum_{\ell =0}^{\infty}\sum_{m=-\ell}^{\ell} {\mathrm{i}}^{\ell}\ 
\frac{1}{3} n\ \delta_{\ell 1}\ Y^{\ast}_{\ell m}\!(\Theta ,\Phi)\ 
Y_{\ell m}\!(\theta ,\phi),   \]
\item[term II] contains all spherical harmonics 
\[ {\mathrm{i}} {\mathbf{n\cdot e}}\ (1-\xi)\ 
{\mathrm{e}}^{ {\mathrm{i}} {\mathbf{n\cdot e}} \xi} = 4\pi\ 
\sum_{\ell =0}^{\infty}\sum_{m=-\ell}^{\ell} \sum_{f=-1}^{1}\ 
{\mathrm{i}}^{\ell +1}\ \frac{4\pi}{3} n(1-\xi)\ j_{\ell}(n\xi)\ 
Y^{\ast}_{1 f}(\Theta ,\Phi)\ Y^{\ast}_{\ell m}\!(\Theta ,\Phi)\ 
Y_{1 f}(\theta ,\phi)\ Y_{\ell m}\!(\theta ,\phi) ,\] 
\item[term III] is a purely monopolar contribution 
\[ 1 = 4\pi\ Y_{00}\!(\theta,\phi)\ Y^{\ast}_{00}\!(\Theta,\Phi) = 4\pi\ 
\sum_{\ell =0}^{\infty}\sum_{m=-\ell}^{\ell}\ {\mathrm{i}}^{\ell}\ 
\delta_{\ell 0}\ 
Y^{\ast}_{\ell m}\!(\Theta ,\Phi)\ Y_{\ell m}\!(\theta ,\phi), \]
\item[term IV] contains all spherical harmonics 
\[  \exp \{ {\mathrm{i}} {\mathbf{n\cdot e}}\xi\} = 4\pi\ 
\sum_{\ell =0}^{\infty} 
\sum_{m=-\ell}^{\ell}\ {\mathrm{i}}^{\ell}\ j_{\ell}(n\xi)\ 
Y_{\ell m}^{\ast}\!(\Theta ,\Phi)\ Y_{\ell m}\!(\theta ,\phi). \]
\end{description}

The part of $g_{\ell m}$ produced by the sum of terms I, III and IV is easy 
to calculate: 
\[ \stackrel{(1)}{g}\!\!\!\!_{\ell m} = \frac{4\pi}{(2\pi)^{3/2}}\ 
{\mathrm{i}}^{\ell}\ \int_{-\infty}^{+\infty} {\mathrm{d}}^{3}{\mathbf{n}}\ 
B_{{\mathbf{n}}}\ \left[ \frac{1}{3}n \delta_{\ell 1} +  
\delta_{\ell 0} -j_{\ell}\!(n\xi )\right]\ Y_{\ell m}^{\ast}\!(\Theta ,\Phi) 
.\] 
The part of $g_{\ell m}$ contributed by term II demands a more tedious 
calculation, which results in
\[ \stackrel{(2)}{g}\!\!\!\!_{\ell m} = \frac{4\pi}{(2\pi)^{3/2}}\ 
{\mathrm{i}}^{\ell}\ \int_{-\infty}^{+\infty} {\mathrm{d}}^{3}{\mathbf{n}}\ 
B_{{\mathbf{n}}}\ n(1-\xi)\ \frac{1}{2\ell +1} [ (\ell +1)\ 
j_{\ell +1}\!(n\xi ) -\ell \ j_{\ell -1}\!(n\xi )]\ 
Y_{\ell m}^{\ast}\!(\Theta ,\Phi) .\] 
An outline of the calculation may be found in Appendix \ref{sec-B}. The total 
expression 
$g_{\ell m} = \stackrel{(1)}{g}\!\!\!\!_{\ell m} + 
\stackrel{(2)}g\!\!\!\!_{\ell m}$ is 
   \begin{equation}
       g_{\ell m} = \frac{4\pi}{(2\pi)^{3/2}}\ {\mathrm{i}}^{\ell}\ 
\int_{-\infty}^{+\infty} {\mathrm{d}}^{3}{\mathbf{n}}\ B_{{\mathbf{n}}}\ 
D_{\ell}(n) 
\ Y_{\ell m}^{\ast}\!(\Theta ,\Phi)  ,    \label{eq:glm}
   \end{equation}
where $D_{\ell}(n)$ denotes the combination 
   \begin{equation}
       D_{\ell}(n) \equiv \frac{1}{3} n\ \delta_{\ell 1} + n (1-\xi)\ 
\frac{1}{2\ell +1} [ (\ell +1)\ j_{\ell +1}\!(n\xi) -\ell \ 
j_{\ell -1}\!(n\xi)] + \delta_{\ell 0} - j_{\ell}(n\xi) .     \label{eq:Dl}
   \end{equation}
The terms in $D_{\ell}(n)$ shall be referred to as term A, term B, term C 
and term D, in the order of their appearance 
in (\ref{eq:Dl}). They originate from the terms I - IV introduced earlier, 
respectively. For the three cases $\ell=0$, 
$\ell=1$ and $\ell \geq 2$, $D_{\ell}(n)$ is explicitly given by:
   \begin{eqnarray}
       D_{0}\!(n) &=& n (1-\xi)\ j_{1}\!(n\xi) + 1 - j_{0}\!(n\xi) ,  
\nonumber \\
       D_{1}\!(n) &=& \frac{1}{3} n + n(1-\xi)\ \frac{1}{3} 
[ 2 j_{2}\!(n\xi) - j_{0}\!(n\xi)] - j_{1}\!(n\xi) ,  \label{eq:differentDl} \\
       D_{\ell \geq 2}\!(n) &=& n(1-\xi)\ \frac{1}{2\ell +1} 
[ (\ell +1)\ j_{\ell +1}\!(n\xi) -\ell\ j_{\ell -1}\!(n\xi)] - 
j_{\ell}(n\xi) .\nonumber   
   \end{eqnarray}
Finally, we can write 
   \begin{equation}
      \frac{\delta T}{T}\!({\mathbf{e}}) =\frac{1}{10}\ 
\sum_{\ell =0}^{\infty} \sum_{m=-\ell}^{\ell} [g_{\ell m}\ 
Y_{\ell m}\!(\theta ,\phi) + g_{\ell m}^{\ast}\ 
Y_{\ell m}^{\ast}\!(\theta ,\phi)]      \label{eq:tempanisspher}
   \end{equation}
where the $g_{\ell m}$ are given by (\ref{eq:glm}). 

The calculation of the integral (\ref{eq:SWformula}) requires only the 
knowledge of the gravitational field perturbations 
$h_{ij}$. However, 
for further references, we remind the expression for the growing density 
contrast 
$(\delta \rho/\rho)\!({\mathbf{x}},\eta)$ associated with the solution 
(\ref{eq:firstpertdp}):
  \[   \frac{\delta \rho}{\rho}\!({\mathbf{x}}, \eta) = -\frac{1}{20}\ 
(\eta - \eta_{{\mathrm{m}}})^{2}\ \nabla^{2}\ B({\mathbf{x}}) .\]
The density contrast is a product of a purely $\eta$-dependent part 
$(\eta - \eta_{{\mathrm{m}}})^{2}$ and a purely spatial part
\begin{equation}
 \frac{\delta \rho}{\rho}\!({\mathbf{x}}) \equiv -\frac{1}{20} 
\nabla^{2}\ B({\mathbf{x}}) . \label{eq:densitycontrastspatial}
\end{equation}
Equation (\ref{eq:densitycontrastspatial}) and the decomposition
\begin{equation}
  \frac{\delta \rho}{\rho}\!({\mathbf{x}}) = (2 \pi)^{-3/2} 
\int_{-\infty}^{+\infty} {\mathrm{d}}^{3}{\mathbf{n}} 
\left[\left(\frac{\delta \rho}{\rho} \right)_{{\mathbf{n}}}\ 
{\mathrm{e}}^{{\mathrm{i}} 
{\mathbf{n\cdot x}}} + 
\left(\frac{\delta \rho}{\rho}\right)^{\ast}_{{\mathbf{n}}}\ 
{\mathrm{e}}^{-{\mathrm{i}} {\mathbf{n\cdot x}}} \right]  
\label{eq:densitycontrastfourrier}
\end{equation}
establish the link between the Fourier components of the density variation and
the gravitational field perturbation: 
\begin{equation}
  \left(\frac{\delta \rho}{\rho}\right)_{{\mathbf{n}}} = 
\frac{1}{20}n^{2}\cdot B_{{\mathbf{n}}} . \label{eq:fourcomprel}
\end{equation}

\section{Statistical assumptions and multipole moments}
\label{sec-SECOND}

Formulae (\ref{eq:tempanisfirst}), (\ref{eq:tempanisG}) give a temperature 
variation over the sky, which is produced by a single and 
deterministic, even if very complicated, perturbed configuration of matter 
density and gravitational field. We believe, however, 
that primordial cosmological perturbations may have had a quantum-mechanical 
origin. To this end, the quantities $h_{ij}$ 
and $\delta T/T$ are quantum-mechanical operators, and the extraction of the 
observable information should proceed through 
the calculation of quantum-mechanical expectation values. Later on, we shall 
go into some details of the 
quantum-mechanical generation of cosmological perturbations. However, in this 
Section, we want to keep the discussion at 
a phenomenological and as simple as possible level. 

We make a simplifying assumption that the quantities $B_{{\mathbf{n}}}$ are 
classical random functions. Since 
$B_{{\mathbf{n}}}$ appears in the expression of $\delta T/T$, the temperature 
variation is also a random function. The 
angular brackets $<...>$ will denote the ensemble average. A single 
statistical hypothesis that we shall need in the 
subsequent calculations is expressed by the equalities  
$\langle B_{{\mathbf{n}}} \rangle = 0$ and
   \begin{equation}
      \langle B_{{\mathbf{n}}}\  B^{\ast}_{{\mathbf{p}}} \rangle = 
|B_{n}|^{2}\ \delta^{3}\!({\mathbf{n}} - {\mathbf{p}}) = 
\langle B_{{\mathbf{p}}}\  B^{\ast}_{{\mathbf{n}}}  \rangle ,
\ \ \ \ \ \ 
      \langle B_{{\mathbf{n}}}\ B_{{\mathbf{p}}}  \rangle = 0 = 
\langle B^{\ast}_{{\mathbf{n}}}\ B^{\ast}_{{\mathbf{p}}} \rangle. 
\label{eq:means}
   \end{equation}
The quantity $|B_{n}|^{2}$ characterises the power spectrum of 
$B\!({\mathbf{x}})$. Indeed, by multiplying 
(\ref{eq:Bfourrier}) with itself and using  (\ref{eq:means}), one finds
\begin{equation}
  \langle B\!({\mathbf{x}})\ B\!({\mathbf{x}}) \rangle = \pi^{-2}\ 
\int_{0}^{\infty} \frac{{\mathrm{d}}n}{n}\ n^{3}\ |B_{n}|^{2} . 
\label{eq:powerB}
\end{equation}

The quantity we need to calculate is the correlation function 
\[ \left\langle \frac{\delta T}{T}\!({\mathbf{e_{1}}})\ 
\frac{\delta T}{T}\!({\mathbf{e_{2}}}) \right\rangle . \]
The vectors ${\mathbf{e_{1}}},\ {\mathbf{e_{2}}}$ refer to two different 
directions of observation, separated by the 
angle $\gamma_{12}$. Using the product of two expressions 
(\ref{eq:tempanisspher}), one
finds 
       \begin{eqnarray}
          \langle \frac{\delta T}{T}({\mathbf{e_{1}}}) \ 
\frac{\delta T}{T}({\mathbf{e_{2}}}) \rangle &=& \frac{1}{100}
\sum_{\ell=0}^{\infty}\sum_{m=-\ell}^{\ell} \sum_{\ell'=0}^{\infty}
\sum_{m'=-\ell'}^{\ell'} \left[  \langle  g_{\ell m}\ 
g^{\ast}_{\ell' m'}\rangle\ Y_{\ell m}\!(\theta_{1},\phi_{1})\ 
Y^{\ast}_{\ell' m'}\!(\theta_{2},\phi_{2}) +  \langle  g_{\ell m}\ 
g_{\ell' m'}\rangle\ Y_{\ell m}\!(\theta_{1},\phi_{1})\ 
Y_{\ell' m'}(\theta_{2},\phi_{2}) \right. \nonumber \\
           &+& \left. \langle  g^{\ast}_{\ell m}\ g^{\ast}_{\ell' m'}\rangle\ 
Y^{\ast}_{\ell m}\!(\theta_{1},\phi_{1})\ 
Y^{\ast}_{\ell' m'}\!(\theta_{2},\phi_{2}) + \langle  g^{\ast}_{\ell m}\ 
g_{\ell' m'}\rangle \ Y^{\ast}_{\ell m}\!(\theta_{1},\phi_{1})\ 
Y_{\ell' m'}\!(\theta_{2},\phi_{2}) \right].       \label{eq:step1}
       \end{eqnarray}
Using (\ref{eq:glm}), (\ref{eq:means}), and the orthogonality condition for 
spherical 
harmonics (see, for example, formula (3.55) of Jackson \shortcite{Jack}), one 
derives
\[ \langle  g_{\ell m}\ g_{\ell' m'}\rangle \ =0=\ \langle g^{\ast}_{\ell m}\ 
g^{\ast}_{\ell' m'}\rangle \]
and
        \begin{equation}
            \langle  g_{\ell m}\ g^{\ast}_{\ell' m'}\rangle \ =\ \langle  
g^{\ast}_{\ell m}\ g_{\ell' m'}\rangle =\ 2\pi^{-1}\ (-1)^{\ell'}\ 
{\mathrm{i}}^{\ell + \ell'}\ \left[ \int_{0}^{\infty} 
\frac{{\mathrm{d}}n}{n}\ n^{3}\ 
|B_{n}|^{2}\  D_{\ell}(n)\ D_{\ell'}(n) \right]\ \delta_{\ell \ell'}\ 
\delta_{m m'}  .   \label{eq:step2}
        \end{equation}

The substitution of (\ref{eq:step2}) in (\ref{eq:step1}) and the application 
of the addition theorem for spherical 
harmonics (see, for example, formula (3.62) of Jackson \shortcite{Jack}), 
brings the 
correlation function to the form
     \[    \langle \frac{\delta T}{T}({\mathbf{e_{1}}}) \ 
\frac{\delta T}{T}({\mathbf{e_{2}}}) \rangle = \sum_{\ell =0}^{\infty} 
\frac{2\ell +1}{4\pi} \  C_{\ell} \ P_{\ell}(\cos \gamma_{1 2}) ,\]
where the multipole moments $C_{\ell}$ are given by the expression
     \begin{equation}
        C_{\ell} = \frac{4\pi^{-1}}{100}\int_{0}^{\infty} 
\frac{{\mathrm{d}}n}{n} \ 
n^{3}\ |B_{n}|^{2}\ D_{\ell}^{2}\!(n). \label{eq:firstCl}
     \end{equation}
Thus, the multipole moments are fully determined by the dimensionless power 
spectrum $n^3 |B_{n}|^{2}$ and the square of 
the known function (\ref{eq:Dl}). 

There is no physical reason why the function $|B_{n}|^{2}$ should be a power 
law function of the wavenumber $n$, let 
alone to have a fixed spectral index in a broad interval of $n$. For 
cosmological perturbations of quantum mechanical 
origin, a power law spectrum in terms of $n$ is generated by a power law 
(in terms of $\eta$) scale factor $a_{{\mathrm{i}}}(\eta)$ 
describing the very early Universe. But, every deviation of 
$a_{{\mathrm{i}}}(\eta)$ 
from a power law in $\eta$ produces 
a deviation of the generated spectrum from a power law in $n$ \cite{GriSo}. 
Virtually any observed $|B_{n}|^{2}$ as a 
function 
of $n$ can be attributed to a specially chosen generating function 
$a_{{\mathrm{i}}}(\eta)$. Had we known $|B_{n}|^{2}$ 
\textit{a priori}, we would find the other cosmological parameters by 
comparing $C_{\ell}$ with observations. If we knew 
the other cosmological parameters with certainty, we would deduce from 
observations $|B_{n}|^{2}$ and $a_{{\mathrm{i}}}(\eta)$. In 
reality, however, almost everything should be found from observations. To 
simplify the problem, we shall follow the common 
tradition and postulate the power law dependence 
     \begin{equation}
|B_{n}|^{2} = K\!(\beta)\ n^{2\beta +1}   , \label{eq:powerspectrumform}
     \end{equation}
where the coefficient $K(\beta)$ is a constant and $\beta$ is a number. In 
the theory of quantum mechanical 
(superadiabatic, parametric) generation of cosmological perturbations, 
$\beta$ enters the power law 
scale factor $a_{{\mathrm{i}}}(\eta) = l_{0} |\eta|^{1+\beta}$, with 
$\beta < -1$. But 
for the 
purposes of this presentation, $\beta$ can be regarded as a phenomenological 
parameter, describing the spectral index. 
The constant $K\!(\beta)$ is calculable from the quantum normalisation, but 
for the purposes of this presentation it 
suffices to consider it simply as a number. 

It is customary to write the power spectrum of 
$(\delta \rho/\rho)\!({\mathbf{x}})$ in the form 
\[ {\mathcal{P}}\!(k) \propto k^{{\mathrm{n}}} ,\]
where $k$ is our wavenumber $n$. We denote the spectral index by a 
roman-style n, in contrast to the wavenumber, which is 
always 
denoted by an italic $n$. It is easy to relate n with $\beta$. Indeed, 
using (\ref{eq:densitycontrastfourrier}), 
(\ref{eq:means}) and (\ref{eq:fourcomprel}), one finds 
 \[ \left\langle \left( \frac{\delta \rho}{\rho} \right)\!({\mathbf{x}})\ 
\left( \frac{\delta \rho}{\rho} \right)\!({\mathbf{x}}) \right\rangle = 
\pi^{-2}\ \int_{0}^{\infty} \frac{{\mathrm{d}}n}{n}\ n^{3}\ 
(n^{4}\ |B_{n}|^{2}) .\] 
According to our notation, ${\mathcal{P}}\!(k)$ is $(n^{4}\ |B_{n}|^{2})$, so 
the spectral index n is related to 
$\beta$ by the relation ${\mathrm{n}}= 2\beta +5$. The flat spectrum is 
defined by the requirement that the dimensionless 
r.m.s. metric perturbation $n^{3/2}|B_{n}|$ (per logarithmic interval of $n$) 
does not depend on $n$.
This gives $\beta = -2$ or ${\mathrm{n}} = 1$.

The substitution of (\ref{eq:powerspectrumform}) in (\ref{eq:firstCl}) leads 
to 
      \begin{equation}
         C_{\ell} = \frac{4\pi^{-1}}{100} K(\beta)\ \int_{0}^{\infty} 
n^{2(\beta +2)}\ D^{2}_{\ell}\!(n)\ \frac{{\mathrm{d}}n}{n} .      
\label{eq:temporaryCl}
      \end{equation}
This formula reduces, for $\ell \geq 2$, to an expression similar to equation 
(17) of Abbott \& Wise \shortcite{AbWi} (given there without 
derivation).

\section{Formula (1) and the limits of its applicability}
\label{sec-THIRD}

Formula (\ref{eq:one}) arises from the (ill-justified) assumption that all 
terms in (\ref{eq:tempanisfirst}) can be 
neglected with respect to the last one, term IV. Accordingly, only the last 
term in (\ref{eq:Dl}), term D, is being 
retained. Formula (\ref{eq:temporaryCl}) is, thus, truncated to
   \begin{equation}
      {\mathcal{C}}_{\ell} =  \frac{4\pi^{-1}}{100}\ K\!(\beta)\ 
\int_{0}^{\infty} \left[-j_{\ell}(n\xi)\right]^{2}\ n^{2(\beta +2)}\  
\frac{{\mathrm{d}}n}{n} . \label{eq:mutilatedCl}
   \end{equation} 
We use a calligraphic ${\mathcal{C}}_{\ell}$ to denote the multipole moments 
resulting from this assumption, in 
contrast to the $C_{\ell}$ that follow from the correct formula 
(\ref{eq:temporaryCl}), where all terms are being 
retained. Using standard relations (see paragraph 10.1.1 of Handbook of 
Mathematical Functions \shortcite{AbSte} and 
formula (6.574.2) of Gradshteyn \& Ryzhik \shortcite{GraRy}), the 
integral (\ref{eq:mutilatedCl}) is analytically calculated 
\cite{Peeb,BondEf}: 
   \begin{equation}
     {\mathcal{C}}_{\ell} =  \frac{2^{2\beta +1}}{25}\ \xi^{-2(\beta +2)}\ 
\frac{\Gamma\!(-2\beta -2)\ \Gamma\!(\ell + \beta +2)}{\left[ 
\Gamma\!(-\beta -\frac{1}{2}) \right]^{2}\ \Gamma\!(\ell -\beta)}\ 
K\!(\beta),\ \ \ {\mathrm{for}}\ \ell>-\beta -2\ {\mathrm{and}}\ \beta < -1 . 
\label{eq:generalmutilated} 
   \end{equation}
For the particular case of the flat spectrum, the multipole moments are found 
by substituting $\beta =-2$ in 
(\ref{eq:generalmutilated}): 
   \begin{equation}
      {\mathcal{C}}_{\ell} =  \frac{(50\pi)^{-1}\ K\!(-2)}{\ell (\ell +1)},\ 
\ \ {\mathrm{for}}\ \ell>0  .      \label{eq:flatmutilated} 
   \end{equation}
This expression is in fact formula (\ref{eq:one}), the $const$ being 
$(50\pi)^{-1}\ K\!(-2)$. Strictly speaking, the 
monopole moment $\ell =0$ is not contained in (\ref{eq:flatmutilated}), but 
the suggested infinity follows from a 
rigorous calculation. One needs to return to (\ref{eq:mutilatedCl}), setting 
$\ell =0$ and $\beta=-2$, to derive 
   \begin{equation}
     {\mathcal{C}}_{0} = \frac{4\pi^{-1}}{100}\ K\!(\beta)\ \int_{0}^{\infty} 
\left[\frac{\sin\!(n\xi)}{(n\xi)}\right]^{2}\ \frac{{\mathrm{d}}n}{n} . 
\label{eq:flatmutilatedC0}
   \end{equation}
In the short wavelength limit $n \rightarrow +\infty$, the integral 
(\ref{eq:flatmutilatedC0}) is well behaved. However, 
in the opposite limit $n \rightarrow 0$, the quantity 
$\sin\!(n \xi)/(n\xi) \approx 1$, and ${\mathcal{C}}_{0}$ is 
determined by the logarithmically divergent expression.
 
The predicted infinity of the monopole moment is not a reflection of some 
inherent property of the flat spectrum, or some 
artificial `gauge problem', as many think, but is strictly the result of 
using (\ref{eq:mutilatedCl}) instead of the 
correct formula (\ref{eq:temporaryCl}). As was already mentioned in the 
Introduction, formula (\ref{eq:flatmutilated}) 
also predicts that the dipole, $\ell =1$, and quadrupole, $\ell=2$, moments 
are of the same order of magnitude, in 
conflict with observations.

It is instructive to explore the integrand of formula (\ref{eq:temporaryCl}) 
separately in the long-wavelength 
$n \ll n_{{\mathrm{H}}}$ and short-wavelength $n \gg n_{{\mathrm{H}}}$ 
regimes. This analysis will 
show when and what kind of dangers one 
can expect. It will also demonstrate some dramatic differences in the 
asymptotic behaviour of the integrands in 
(\ref{eq:temporaryCl}) and (\ref{eq:mutilatedCl}). The most spectacular 
modifications occur in the long-wavelength 
behaviour of the monopole moment and short-wavelength behaviour of the dipole 
moment, but all multipoles are seriously 
affected.
For this analysis we shall use a strictly power-law spectrum 
(\ref{eq:powerspectrumform}), bearing in mind, 
though, that the transfer function effectively suppresses the power in short 
waves, so that a real spectrum is 
turned down at sufficiently large $n$'s. It is convenient to separately 
examine the cases $\ell =0$, $\ell =1$ and 
$\ell \geq 2$ in each regime. 

{\textbf{The long-wavelength contribution to}} ${\mathbf{C_{0}}}$: The 
expression of $D_{0}\!(n)$ is given by the first 
line in 
(\ref{eq:differentDl}). The participating terms are B, C and D. We shall show 
that the presence of the purely monopolar 
term C cancels out the infinity arising due to term D, making the monopole 
moment finite and small. Expanding the Bessel 
functions $j_{0}\!(n\xi)$ and $j_{1}\!(n\xi)$ up to $(n\xi)^{2}$ and $(n\xi)$ 
respectively, one derives the approximate 
expression for $D_{0}\!(n)$ in the long wavelength limit:
 \[ D_{0}\!(n) \approx \frac{1}{3}\ n^{2}\ (1-\xi)\xi + 1 -1 + \frac{1}{6}\ 
n^{2}\ \xi^{2} = \frac{1}{3}\left( 1- \frac{1}{2}\xi \right)\xi\ n^{2} .\]
Term C, which is $1$, exactly cancels with the leading order term in the 
expansion of term D, which is $-1$. As a result, 
$D_{0}\!(n)$ is of the order $(n\xi)^{2} \ll 1$ and not unity. Formula 
(\ref{eq:temporaryCl}) yields
\begin{equation}
  C_{0} \approx \frac{4\pi^{-1}}{100} K(\beta)\ \frac{1}{9}
\left(1- \frac{1}{2}\xi \right)^{2}\xi^{2}\ \int_{0}^{s} 
\frac{{\mathrm{d}}n}{n}\ n^{2(\beta +4)} . \label{eq:C0long}
\end{equation}
It is clear that the monopole moment $C_{0}$, unlike ${\mathcal{C}}_{0}$, is 
finite in the case of the flat ($\beta =-2$) 
spectrum. The integral (\ref{eq:C0long}) is convergent for all $\beta$'s 
satisfying the constraint 
\begin{equation}
  \beta > -4 .\label{eq:conC0}
\end{equation}
The monopole moment is genuinely divergent in the long-wavelength limit only 
for such `red' spectra, that $\beta$ 
violates (\ref{eq:conC0}).

{\textbf{The long-wavelength contribution to}} ${\mathbf{C_{1}}}$: 
$D_{1}\!(n)$ is given by the second line in 
(\ref{eq:differentDl}). 
The participating terms are A, B and D. The purely dipolar term A combines 
with terms B and D in such a way, that the 
dipole moment in this regime is suppressed, as compared with the monopole and 
quadrupole. With the help of appropriate 
expansions of the Bessel functions one derives
 \[ D_{1}\!(n) \approx \frac{1}{3}n -\frac{1}{3}\ n(1-\xi) +\frac{3}{30}
(1-\xi)\xi^{2}\ n^{3} -\frac{1}{3}\ n\xi +\frac{1}{30}n^{3}\xi^{3} = 
\frac{1}{30}\ (3-2\xi)\xi^{2}\ n^{3} . \]
Term A combines with the leading order terms in B and D. As a result, 
$D_{1}\!(n)$ is of the order $(n\xi)^{3}$ and not 
$(n\xi)$, as would be the case if only term D were retained. In agreement 
with Grishchuk \& Zel'dovich \shortcite{GriZe}, $D_{1}\!(n)$ has one extra 
power of $n$ (which is a small number for long waves) as compared with 
$D_{0}\!(n)$ and $D_{2}\!(n)$  (for $D_{2}\!(n)$ see (\ref{eq:D>2long}) 
below). Formula (\ref{eq:temporaryCl}) yields
\begin{equation}
  C_{1} \approx \frac{4\pi^{-1}}{100} K(\beta)\ \frac{1}{900}
( 3-2\xi)^{2}\xi^{4}\ \int_{0}^{s} \frac{{\mathrm{d}}n}{n}\ n^{2(\beta +5)} .
\label{eq:C1long}
\end{equation} 

{\textbf{The long-wavelength contribution to}} ${\mathbf{C_{\ell \geq 2}}}$: 
The expression for $D_{\ell \geq 2}\!(n)$ is 
given by the 
third line in (\ref{eq:differentDl}). The participating terms are B and D. 
Expanding the Bessel 
functions, one finds
\begin{equation}
  D_{\ell \geq 2}\!(n) \approx 
-\frac{[\xi + \ell (1-\xi)]\xi^{\ell -1}}{1\cdot3\cdot \ldots \cdot (2\ell +1)}
\ n^{\ell} \label{eq:D>2long}
\end{equation}
and formula (\ref{eq:temporaryCl}) yields
\begin{equation}
  C_{\ell \geq 2} \approx \frac{4\pi^{-1}}{100} K(\beta)\ 
\left[-\frac{[\xi + \ell (1-\xi)]\xi^{\ell -1}}{1\cdot3\cdot \ldots \cdot (2\ell +1)} 
\right]^{2}\ \int_{0}^{s} 
\frac{{\mathrm{d}}n}{n}\ n^{2(\beta +2 +\ell)} .
\label{eq:C>2long}
\end{equation}
As expected \cite{GriZe,LPG2}, $C_{0}$, given by (\ref{eq:C0long}), and 
$C_{2}$, given by (\ref{eq:C>2long}) for 
$\ell =2$, have exactly the same behaviour in the limit $n \rightarrow 0$. 
The integral (\ref{eq:C>2long}) converges if 
$\beta$ satisfies the constraint $\beta > -2 -\ell$. Not surprisingly, this 
constraint for $\ell =2$ reduces to 
(\ref{eq:conC0}). The same constraints on $\beta$ follow from the requirement 
that the quadrupole moment produced by 
gravitational waves does not diverge in the long-wavelength regime 
\cite{LPG1}.

We now examine formula (\ref{eq:temporaryCl}) in the short wavelength limit. 
In this analysis, we ignore the fact that 
the primordial power-law spectrum is bent down by the modulating (transfer) 
function and, in any case, the integration 
should be terminated at some large $n= n_{{\mathrm{max}}}$ due to the 
invalidation of 
the linear approximation and the beginning of 
a non-linear regime for density perturbations. The purpose of this analysis 
is to examine how the integrals 
(\ref{eq:temporaryCl}) behave as the upper limit of integration increases. 
The proper inclusion of the modulating 
function will be performed in the next Section.  

{\textbf{The short-wavelength contribution to}} ${\mathbf{C_{0}}}$: In the 
limit of large $n$'s, the asymptotic behaviour 
of 
$D_{0}\!(n)$ (first line in (\ref{eq:differentDl}) ) is dominated by the 
purely monopolar term C and the leading order 
expansion of term B:
\[ D_{0}\!(n) \approx 1\ -\ \frac{1-\xi}{\xi}\ \cos(n\xi) .\]
The short-wavelength contribution to $C_{0}$ is given by 
\begin{equation}
  C_{0} \approx \frac{4\pi^{-1}}{100} K(\beta)\ \int_{s}^{+\infty} 
\frac{{\mathrm{d}}n}{n}\ \left[ 1 - 2\frac{1-\xi}{\xi}\ \cos(n\xi ) \right] 
\ n^{2(\beta +2)} .\label{eq:naiveC0short}
\end{equation}
This integral would be logarithmically divergent in the case $\beta = -2$, 
and power-law divergent for 
$\beta > -2$, if one could extend the integration up to infinity. The 
short-wavelength behaviour of $C_{0}$ is 
drastically different from the one of ${\mathcal{C}}_{0}$, 
given by expression (\ref{eq:flatmutilatedC0}). 

{\textbf{The short-wavelength contribution to}} ${\mathbf{C_{1}}}$: The 
asymptotic behaviour of $D_{1}\!(n)$ is dominated 
by the purely dipolar term A:
 \[  D_{1}\!(n) \approx \frac{1}{3}\ n .\]
As expected, $D_{1}\!(n)$ has one extra power of $n$ (a large number for 
short waves), as compared with $D_{0}\!(n)$ and 
$D_{2}\!(n)$ 
(for $D_{2}\!(n)$ see formula (\ref{eq:D>2short}) below). This makes the 
dipole moment much more sensitive to short 
waves, in comparison with the monopole and quadrupole, as it should. The 
short wavelength contribution to $C_{1}$ is
 \[ C_{1} \approx \frac{4\pi^{-1}}{100} K(\beta)\ \frac{1}{9}\ 
\int_{s}^{+\infty} \frac{{\mathrm{d}}n}{n}\ n^{2(\beta +3)} .\]
The integrand has two extra powers of $n$, as compared with $C_{0}$ and 
$C_{2}$.

{\textbf{The short-wavelength contribution to}} ${\mathbf{C_{\ell \geq 2}}}$: 
The asymptotic behaviour of 
$D_{\ell \geq 2}\!(n)$ is dominated by term B:
\begin{equation}
  D_{\ell \geq 2}\!(n) \approx \frac{1-\xi}{\xi}\ \cos(n\xi) .
\label{eq:D>2short}
\end{equation}
(Strictly speaking, this expression is valid for even $\ell$'s; for odd 
$\ell$'s, the $\cos$ is replaced by the $\sin$.) 
The short wavelength contribution to $C_{\ell \geq 2}$ is given by 
\begin{equation}
  C_{\ell \geq 2} \approx \frac{4\pi^{-1}}{100} K(\beta)\ \left( 
\frac{1-\xi}{\xi}\right)^{2}\ \int_{s}^{+\infty} \frac{{\mathrm{d}}n}{n}\ 
\frac{1 + \cos(2n\xi)}{2}\ n^{2(\beta +2)} .\label{eq:naiveC>2short}
\end{equation}
The integrand of (\ref{eq:naiveC>2short}) is quite similar to the integrand 
of (\ref{eq:naiveC0short}).

We have demonstrated the importance of the terms A, B and C in the derivation 
of the expected CMB multipoles. In the 
short wavelength regime, term D is always sub-dominant in $D_{\ell}(n)$, 
while it is the sole term retained in formula 
(\ref{eq:mutilatedCl}) for ${\mathcal{C}}_{\ell}$. To illustrate the 
difference between the $C_{\ell}$ and 
${\mathcal{C}}_{\ell}$ distributions, we perform an explicit numerical 
calculation for the flat spectrum $\beta = -2$. 
The integral (\ref{eq:temporaryCl}) is taken up to 
$n_{{\mathrm{max}}} = 1260$. It is 
believed that the linear approximation is 
valid for wavelengths up to 100 times shorter than the Hubble radius (see, 
for example, Padmanabhan \shortcite{Padma}) which gives 
$n_{{\mathrm{max}}} \sim 100\cdot n_{{\mathrm{H}}} = 400\pi$. The value of 
$\xi$ is determined by 
$z_{{\mathrm{dec}}}$ (see equation (\ref{eq:redxi})), 
which we take $z_{{\mathrm{dec}}}=1000$. 

\begin{figure}
\centerline{\epsfxsize=10cm \epsfbox{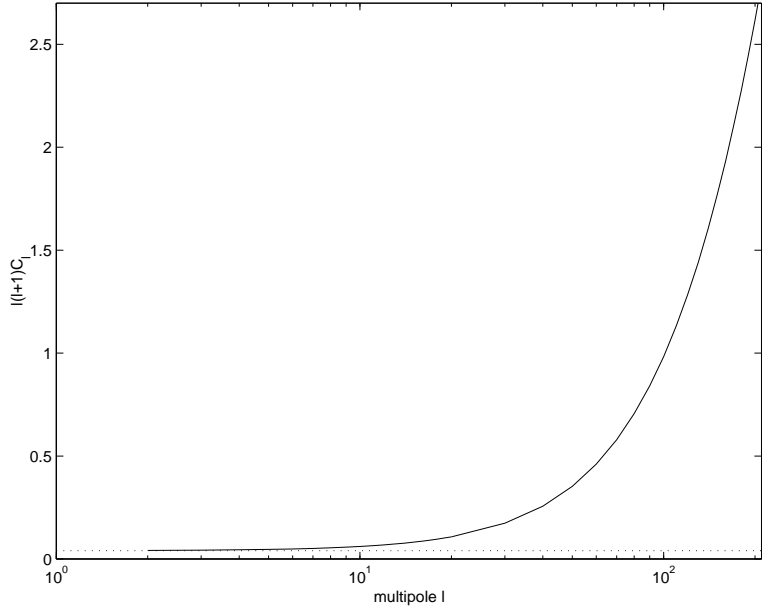}}
\caption{The multipole distributions as described by formula 
(\ref{eq:temporaryCl}) (solid line) and formula 
(\ref{eq:mutilatedCl}) (dotted line) for the flat primordial spectrum 
$\beta= -2$ (${\mathrm{n}} =1$).} 
\label{Beffect}
\end{figure}

In figure \ref{Beffect}, we plot the quantity $\ell (\ell + 1) C_{\ell}$ as a 
function of $\ell$. (For concreteness, we 
take $K\!(-2)=2 \pi$.) Formula (\ref{eq:flatmutilated}) predicts a straight 
horizontal line (shown by the dotted line) 
with 
the monopole moment going to infinity. For the dipole moment $\ell =1$, 
formula (\ref{eq:temporaryCl}) gives 
$\left.\ell(\ell+1)C_{\ell}\right|_{\ell=1} \approx 1.4\cdot 10^{4}$, while 
formula (\ref{eq:flatmutilated}) predicts for 
$\ell=1$ (as well as for all other $\ell$'s) 
$\ell(\ell+1){\mathcal{C}}_{\ell} = 4\cdot 10^{-2}$. The ratio 
$C_{1}/{\mathcal{C}}_{1} \approx 3.5\cdot10^{5}$ clearly illustrates the 
dramatic difference in the expected dipole. The 
higher order multipoles also experience a significant deviation from the 
straight line (mostly, due to the inclusion of 
term B) as shown by the solid line. The ratios 
$C_{\ell}/{\mathcal{C}}_{\ell}$ for 
$\ell =$2, 10, 20, 30, 40, 50 are 1.03, 1.52, 2.69, 4.35, 6.42 and 8.82, 
respectively. The higher order multipoles are 
more sensitive to short waves, and therefore the difference between the lines 
is even more prominent. The growth of the 
function $\ell (\ell+1)C_{\ell}$ would be unlimited, if the spectrum could
be continued as a purely power-law spectrum for $n \gg n_{{\mathrm{max}}}$. 
However, in reality, it is terminated because of a change in the spectrum 
caused by the inclusion of the transfer function.
 
\section{The primordial spectrum and the modulating function}
\label{sec-FOURTH}

A smooth spectrum of density perturbations that existed deeply in the 
radiation-dominated era gets strongly modified by 
the time of decoupling. This is the result of evolution of oscillating 
density perturbations (metric and plasma sound waves) through 
the 
transition from the radiation-dominated regime to the matter-dominated 
regime. This phenomenon is usually described with 
the help of the transfer function (see, for example, 
Peebles \shortcite{Pebook}, Efstathiou \shortcite{Efst}, 
Zel'dovich \& Novikov \shortcite{ZelNo}, Doroshkevich \& Schneider 
\shortcite{DoSch}, Hu \& Sugiyama \shortcite{HuSu}) 
and several model transfer functions are 
known in the literature. Their common property is a strong reduction of the 
short-wavelength power, starting from some 
large critical wavenumber. For concreteness, we shall use the transfer 
function derived in Grishchuk \shortcite{LPG2}. The purpose of 
that paper was to 
derive the power spectrum from first principles and the quantum-mechanical 
approach was consistently used. In a moment, we 
shall briefly outline that derivation in order to relate notations, but the 
final result is not unexpected, so we shall 
present it first. 
The result is the following: instead of the strictly power-law spectrum  
\[ |B_{n}|^{2} = K\!(\beta)\ n^{2\beta +1} \] 
introduced by equation (\ref{eq:powerspectrumform}), the integrals 
(\ref{eq:firstCl}) for the $C_{\ell}$'s should be 
taken with the modified spectrum  
\begin{equation}  
|B_{n}|^{2} = K\!(\beta)\ n^{2\beta +1} M^{2}\!(n\xi_{2}), 
\label{eq:truespectrum}
\end{equation} 
where the modulating (transfer) function $M^{2}\!(n\xi_{2})$ is given by  
\begin{equation} 
M^{2}\!(n\xi_{2}) = \frac{\sin^{2}\!(
\frac{n\xi_{2}}{2\sqrt{3}})}{(\frac{n\xi_{2}}{2\sqrt{3}})^2}  
\label{eq:defMdp} 
\end{equation} 
and $\xi_{2}$ is determined by the red-shift $z_{{\mathrm{eq}}}$ at the 
transition from 
the radiation dominated era to the matter 
dominated era:
\[ \xi_{2} = (1+z_{{\mathrm{eq}}})^{-1/2}.\]
If one does not wish to go into the details of derivation, formula 
(\ref{eq:defMdp}) can be treated as a postulated 
phenomenological transfer function, and formula (\ref{eq:truespectrum}) as a 
postulated resulting spectrum.   

We shall briefly outline the derivation of the power spectrum for density 
perturbations, but the reader interested only in 
applications may skip this paragraph. The theory of quantum-mechanical 
generation of density perturbations 
\cite{LPG2,LPG3} uses the Heisenberg operator for the field 
$h_{ij}(\eta, {\mathbf{x}})$:  
     \begin{equation}
         h_{ij}(\eta,{\mathbf{x}}) = l_{{\mathrm{Pl}}}
\frac{\sqrt{16\pi}}{(2\pi)^{3/2}}\ 
\int_{-\infty}^{\infty} {\mathrm{d}}^{3}{\mathbf{n}}\ \frac{1}{\sqrt{2n}}\ 
\sum_{s=1}^{2}\stackrel{s}{P}_{ij}\!({\mathbf{n}})\ 
\left[\stackrel{s}{h}_{n}\!(\eta)\ \stackrel{s}{c}_{\mathbf{n}}\!(0)\ 
{\mathrm{e}}^{{\mathrm{i}}{\mathbf{n\cdot x}}} + 
\stackrel{s}{h}_{n}^{\ast}\!(\eta)\ 
\stackrel{s}{c}_{\mathbf{n}}^{\dagger}\!(0)\ 
{\mathrm{e}}^{-{\mathrm{i}}{\mathbf{n\cdot x}}}\right] .
 \label{eq:quantumdp}
     \end{equation}
where the Planck length $l_{{\mathrm{Pl}}}$ comes from the quantum 
normalisation of the 
field to a half of a quantum in each mode, 
while $c_{\mathbf{n}}\!(0)$ and $c_{\mathbf{n}}^{\dagger}\!(0)$ are the usual 
annihilation and creation operators. Two 
polarisation tensors $\stackrel{s}{P}_{ij}\!\!(\mathbf{n})$ ($s = 1, 2$) 
describe, respectively, a scalar and a 
longitudinal component of the field:  
\[ \stackrel{1}{P}_{ij}\!\!({\mathbf{n}})=\delta_{ij}\ ,\ 
\stackrel{2}{P}_{ij}\!\!({\mathbf{n}})= - n_{i}n_{j}/n^{2}. \] 
The functions $\stackrel{s}{h}_{n}\!\!(\eta)$ satisfy the perturbed 
Einstein equations and are being evolved through three 
successive stages of evolution: initial (i), radiation-dominated (e) and 
matter-dominated (m), with the respective scale 
factors  
  \begin{eqnarray} 
      a_{{\mathrm{i}}}(\eta) &=& l_{0} |\eta|^{1+\beta},\ \ \beta < -1,
\ \eta \leq \eta_{1},\ \eta_{1} <0 , \nonumber \\ 
      a_{{\mathrm{e}}}(\eta) &=& l_{0} a_{e}
( \eta - \eta_{{\mathrm{e}}}),\ \ \eta_{1} 
\leq \eta \leq \eta_{2} , \label{eq:scalefactorsfinal} \\ 
      a_{{\mathrm{m}}}(\eta) &=& 2 l_{H} (\eta - \eta_{{\mathrm{m}}})^{2}, 
\ \ \eta_{2} 
\leq \eta \leq \eta_{{\mathrm{R}}}  .\nonumber 
  \end{eqnarray} 
The constants participating in (\ref{eq:scalefactorsfinal}) are connected 
with each other by the continuous joining of 
$a(\eta)$ and $a'(\eta)$ at the transition points. At the (m)-stage, the 
growing solution for 
$\stackrel{s}{h}_{n}\!\!(\eta)$  
is       
\[ \stackrel{1}{h}_{n} = C\!(n)\ ,\ \stackrel{2}{h}_{n}\!(\eta) =
\frac{1}{10}\ C\!(n)\ [n(\eta -\eta_{{\mathrm{m}}})]^{2} ,\]
which allows one to relate this form of the solution with expressions 
(\ref{eq:firstpertdp}), (\ref{eq:Bfourrier}). 
Comparing the quantum-mechanical expectation value of the square of the 
generated field $h_{ij}$ with the ensemble 
averaging used in equation (\ref{eq:powerB}), one obtains 
$|C\!(n)|^{2} = |B_{n}|^{2} n$, where the factor $n$ comes from 
the extra factor $1/\sqrt{2n}$ in the definition (\ref{eq:quantumdp}). The 
perturbed Einstein equations plus the imposed 
initial conditions allow one to find the spectrum $|C\!(n)|^{2}$ from first 
principles, and express this quantity in 
terms of the participating parameters and the fundamental constants (see 
equations (82), (81) in Grishchuk \shortcite{LPG2} and accompanying 
discussion). The resulting function $|C\!(n)|^2$ has the form
\[ |C\!(n)|^{2} = K\!(\beta)\ n^{2\beta +2}\ M^{2}\!(n\xi_{2}),\]
where $M^{2}(n\xi_{2})$ is given by (\ref{eq:defMdp}) and $K\!(\beta)$ is a 
known function of $\beta$. For the specific case of the 
flat spectrum, $\beta = -2$,  
\begin{equation}
K(-2)=2\pi\ (\frac{l_{{\mathrm{Pl}}}}{l_{0}})^{2}. \label{eq:K}
\end{equation} 
The modulating function is an oscillatory function of $n$ and has deep 
minima. This can be explained \cite{GriSi} as the 
eventual 
`desqueezing' of some of the generated modes, and is also related to the 
standing wave pattern of the generated 
perturbations and the phenomenon of the Sakharov oscillations. 

\begin{figure}
\centerline{\epsfxsize=12cm \epsfbox{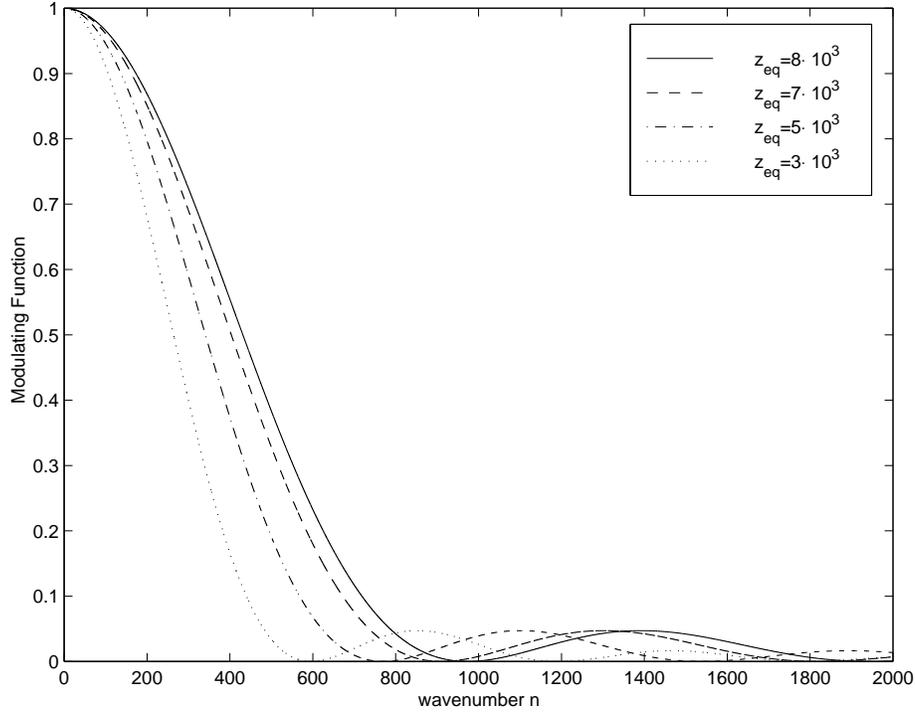}} 
\caption{The modulating function $M^{2}\!(n \xi _{2})$ for different  
values of $z_{{\mathrm{eq}}}$. The wavenumber corresponding to the present-day 
Hubble radius is $4\pi$. For waves longer than and comparable to the Hubble 
radius, the  
modulating function is approximately 1. At wave-numbers around 400 the 
modulating function starts suppressing the primordial power spectrum
significantly.}  
\label{transdp} 
\end{figure}

We shall now work with the function (\ref{eq:truespectrum}), leaving the 
parameters $\beta$, $K\!(\beta)$, $z_{{\mathrm{eq}}}$ and $z_{{\mathrm{dec}}}$ free, to be 
determined by the comparison with observations. In 
figure \ref{transdp}, we plot the modulating function $M^{2}\!(n\xi_{2})$. 

\begin{figure}
\centerline{\epsfxsize=10cm \epsfbox{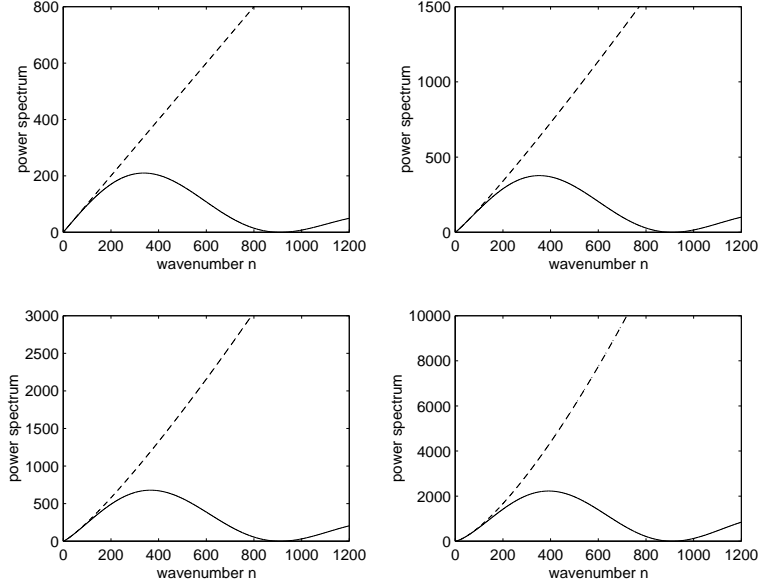}} 
\caption{The effect of the modulating function on different primordial 
spectra $n^{2\beta +5}$. The dashed-dotted line 
refers to the original spectrum $n^{2\beta +5}$, while the solid one to the 
resulting modulated spectrum 
$n^{2\beta +5}\ M^{2}\!(n\xi_{2})$. The spectra shown are: i) top left hand 
side, the flat spectrum $\beta=-2$, ii) 
top right hand side, $\beta=-1.95$, iii) bottom left hand side, $\beta=-1.9$, 
iv) bottom right hand side, $\beta=-1.8$. 
All the modulated spectra were calculated for $z_{{\mathrm{eq}}}=7000$.} 
\label{sakharov} 
\end{figure}

In figure \ref{sakharov}, we present the effect of the modulating function on 
different primordial spectra of the density 
contrast, ${\mathcal{P}}\!(k) = k^{\mathrm{n}}$, which in our notation  
is ${\mathcal{P}}\!(n)= n^{2\beta +5}$. For sufficiently long wavelengths, 
the spectra retain their original form. At 
wave-numbers around 400 the spectra go through the maximum. 

\begin{figure}
\centerline{\epsfxsize=8cm \epsfbox{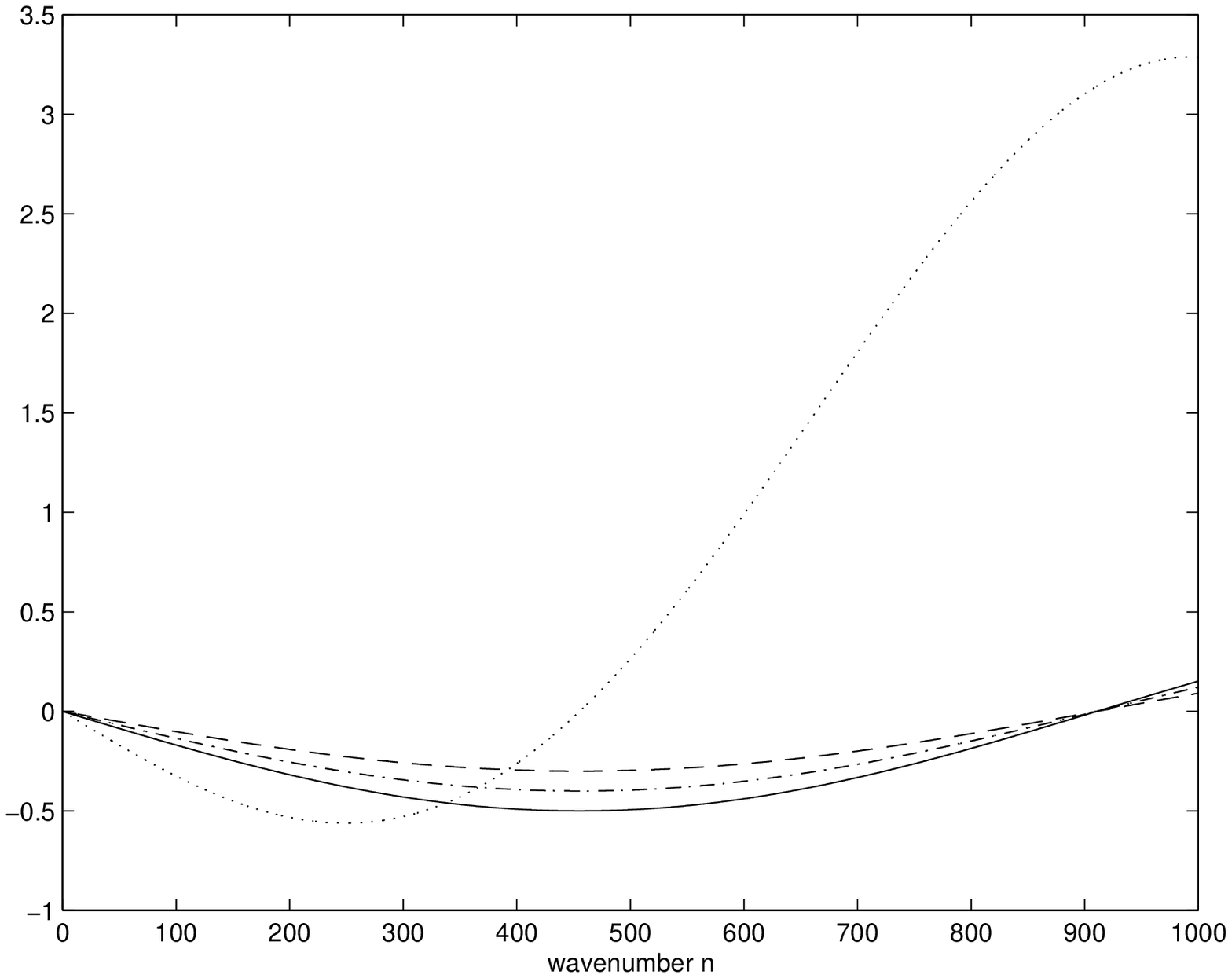}} 
\caption{The dotted line shows the function 
$-[(n\xi _{2})/(2\sqrt{3})]\cdot  
\cos ((n\xi _{2})/(2\sqrt{3}))$, while the  
solid, dashed-dotted and dashed lines show the function  
$[\beta + (3/2)]\cdot \sin ((n\xi _{2})/(2\sqrt{3}))$ for  
$\beta =-2, -1.9,-1.8$ respectively. The red-shift at matter-radiation  
equality is taken $z_{{\mathrm{eq}}}=7000$. The values of the  
wavenumber where the intersection of the curves takes place 
(except the point $n=0$) are the points where the  
modulated power spectrum $n^{2\beta +5}\ M^{2}\!(n\xi_{2})$ has its 
first maximum. For increasing $\beta $, the first maximum occurs at  
larger values of $n$.} 
\label{maximumdp} 
\end{figure}

The maxima of the $n^{2\beta +5}\ M^{2}\!(n\xi_{2})$ spectrum, that we present 
in 
figure \ref{sakharov}, appear at the 
wave-numbers which satisfy the equation  
\[ \left( \beta + \frac{3}{2}\right)\ \sin\!(\frac{n\xi_{2}}{2\sqrt{3}}) =  
-\frac{n\xi_{2}}{2\sqrt{3}}\ \cos\!(\frac{n\xi_{2}}{2\sqrt{3}}).\]  
The increase of $\beta$ displaces the maxima towards larger $n$. As an 
example, we present in figure \ref{maximumdp} the 
displacement of the first maximum towards larger values of $n$, when one 
keeps $z_{{\mathrm{eq}}}=7000$ and increases $\beta$. The 
dotted line represents the right hand side of the equation (which is 
$\beta$-independent), while the solid, dashed-dotted 
and dashed lines represent the left hand side for $\beta=-2\ ,-1.9,\ -1.8$. 
The position of the first maximum in the 
power spectrum is directly related to the position of the first peak in the 
$\ell $-space, as will be discussed below. 

We shall now calculate numerically the multipole moments following from the 
formula  
  \begin{equation}
    C_{\ell} = \frac{4\pi^{-1}}{100} K\!(\beta)\ \int_{0}^{\infty} 
\frac{{\mathrm{d}}n}{n}\ M^{2}\!(n\xi_{2})\ n^{2(\beta +2)}\ 
D_{\ell}^{2}\!(n)    ,  
\label{eq:Clfinal}
  \end{equation}
which replaces formula (\ref{eq:temporaryCl}).

\section{The CMBR anisotropy distributions caused by density perturbations} 
\label{sec-FIFTH} 

\begin{figure}
\centerline{\epsfxsize=10cm \epsfbox{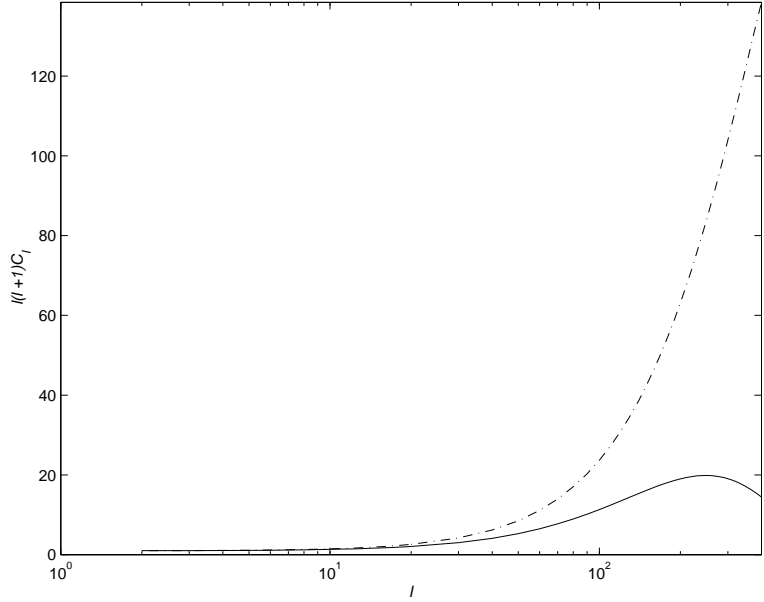}} 
\caption{The quantity $\ell(\ell+1)C_{\ell}$ in the absence (dashed-dotted 
line) and the presence (solid line) of the  
modulating function.}  
\label{sakeffect} 
\end{figure}

The modulating function makes the power spectrum oscillatory and dramatically 
suppresses the power at short waves 
$n\xi_{2} \gg 1$, roughly, in proportion to $n^{-2}$.  
In figure \ref{sakeffect}, we present a typical graph of 
$\ell(\ell+1)C_{\ell}$ calculated with the modulating function. 
The chosen parameters are $z_{{\mathrm{dec}}}=1000$, $z_{{\mathrm{eq}}}=7000$ 
and $\beta=-2$. The 
integration in formula (\ref{eq:Clfinal}) 
is carried out up to $n_{{\mathrm{max}}} =1260$ and the quantity 
$2(2+1)C_{2}$ (for the 
quadrupole, $\ell =2$) is normalised to 1. 
The 
dashed-dotted line shows the result without including the modulating 
function. The solid line clearly shows the first 
peak formed due to the inclusion of the modulating function.  
 
Before proceeding to other choices of the parameters, it is instructive to 
explore the integrand 
$n^{2\beta+3}\ M^{2}\!(n\xi_{2}) D^{2}_{\ell}\!(n)$ in formula 
(\ref{eq:Clfinal}). We shall do this for various $\ell$'s 
and at different intervals of integration over $n$. We shall present the 
graphs of the integrand, which allow one to 
evaluate visually the surface area under the graphs and, hence, the 
contributions to the integrals accumulated at 
different intervals of $n$.

\begin{figure}
\centerline{\epsfxsize=10cm \epsfbox{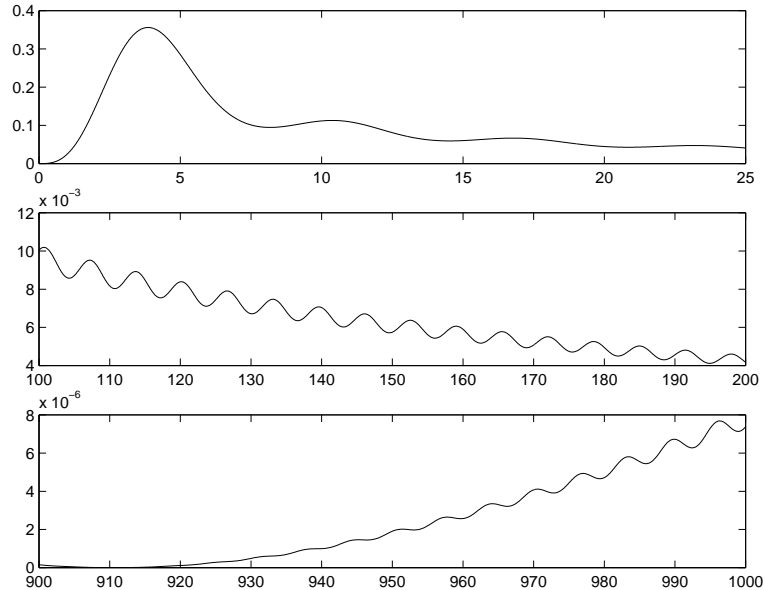}} 
\caption{The integrand of the monopole moment $\ell=0$ versus the  
wavenumber $n$. The values of the parameters are $\beta =-2$, 
$z_{{\mathrm{eq}}}=7000$,  
$z_{{\mathrm{dec}}}=1000$.} 
\label{integrandc0} 
\end{figure}

$\ell =0$: In figure \ref{integrandc0}, we plot the integrand of $C_{0}$ at 
different intervals of $n$. A deep 
minimum of the integrand caused by the first zero of the modulating function 
is clearly seen. It is obvious from the 
graph that the integral is mostly accumulated at small $n$'s, up to 
$n \approx 12$ or so. Thus, the dominant contribution 
to $C_{0}$ comes from wavelengths larger than and comparable to the Hubble 
radius. 

\begin{figure}
\centerline{\epsfxsize=10cm \epsfbox{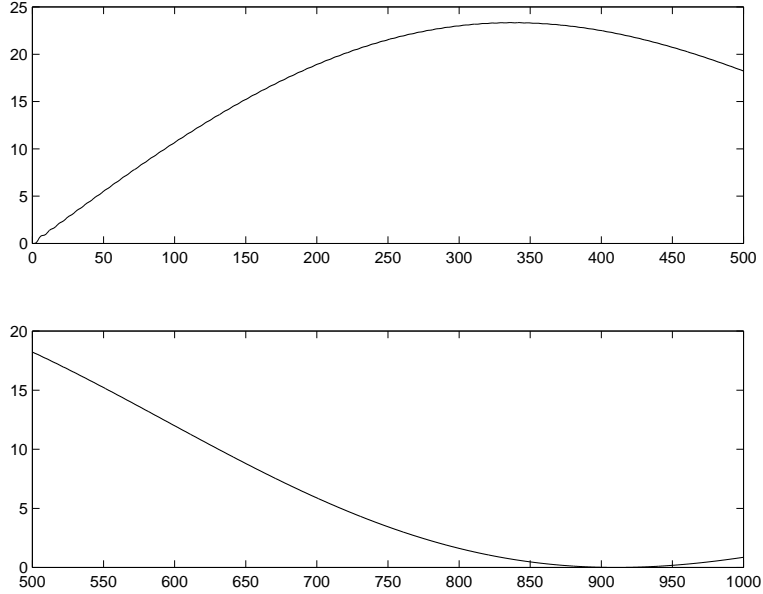}} 
\caption{The integrand of the dipole moment $\ell =1$ versus the  
wavenumber $n$. The values of the parameters are again $\beta =-2$,  
$z_{{\mathrm{eq}}}=7000$, $z_{{\mathrm{dec}}}=1000$.} 
\label{integrandc1} 
\end{figure}

$\ell =1$: In figure \ref{integrandc1}, we plot the integrand 
of $C_{1}$. The modulating function suppresses the short-wavelength power, 
but still leaves the dipole moment 
$C_{1}$ much more sensitive, than $C_0$ and $C_2$, to short waves. The 
remarkable fact is that the main contribution 
comes from wavelengths about 
20--40 times shorter than the Hubble radius.
For the conventional values of the Hubble parameter, this corresponds 
to scales around (100 - 200) Mpc., not much larger and not much smaller.
The most 
important term in the decomposition of $D_{1}\!(n)$ (see formula 
(\ref{eq:differentDl}) ), is term A. It is obvious from 
the graph that the numerical value of the dipole is much greater than the 
value of the monopole and (as we shall show 
below) the $\ell \geq 2$ multipoles. 

\begin{figure}
\centerline{\epsfxsize=12cm \epsfbox{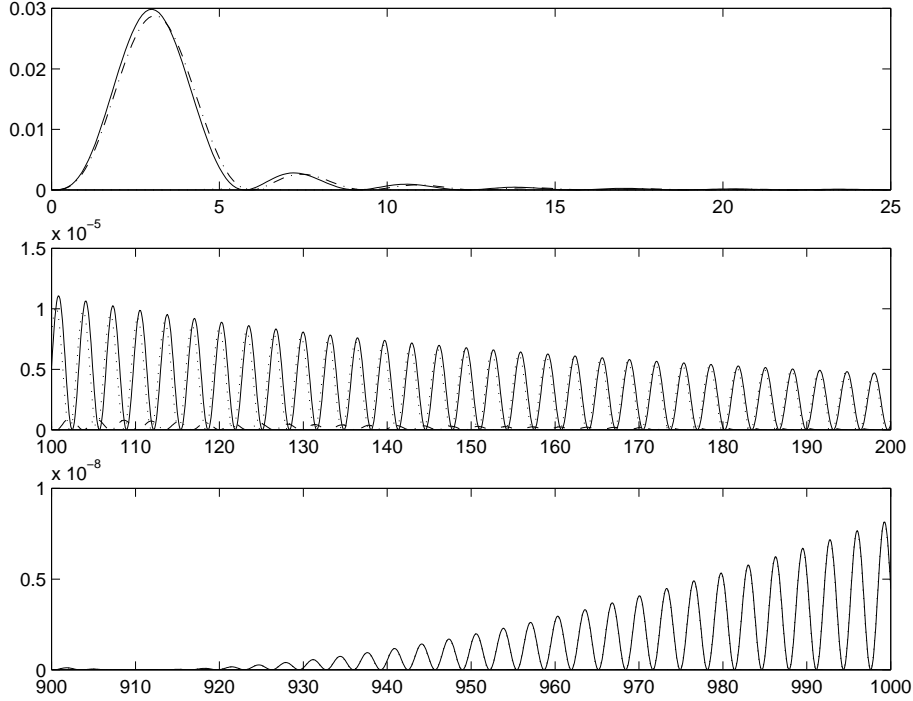}} 
\caption{This plot shows by a solid line the integrand for the quadrupole 
moment, $\ell =2$, at different wavenumber 
intervals. We also display the integrand that results after retaining only 
term D in $D_{2}(n)$ (dashed-dotted line), and 
the integrand that results after retaining only term B (dotted line). For 
long wavelengths, term D clearly dominates over 
term B, while for short wavelengths the situation is reversed. At 
$n \sim 900$, we observe a zero caused by the first 
zero in the modulating function. The modulating function and the Bessel 
functions combine in such a manner, that the area 
under the curve at long and intermediate wavelengths is much greater than the 
area at short wavelengths, where term B 
dominates. The main contribution to $C_{2}$ is provided by large scales.} 
\label{integrandc2dp} 
\end{figure}

\begin{figure}
\centerline{\epsfxsize=12cm \epsfbox{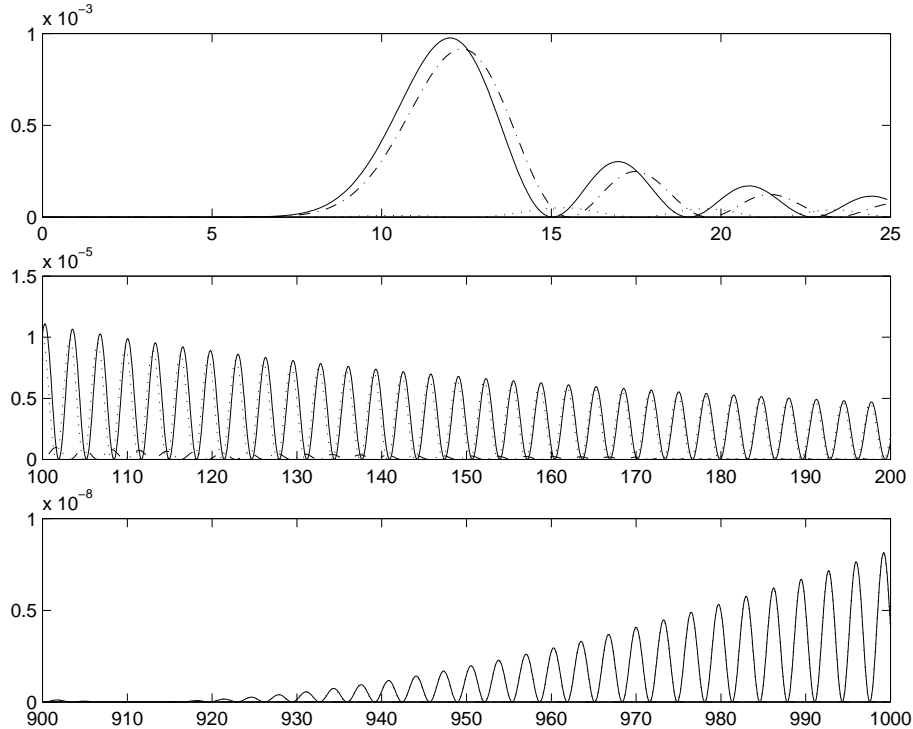}} 
\caption{The same as in figure \ref{integrandc2dp}, for the multipole 
$\ell =10$. Although the role of term B is now 
more important than in the case of the quadrupole, the main conclusions of 
the $\ell =2$ case still apply to $\ell=10$.} 
\label{integrandc10dp} 
\end{figure}

\begin{figure}
\centerline{\epsfxsize=12cm \epsfbox{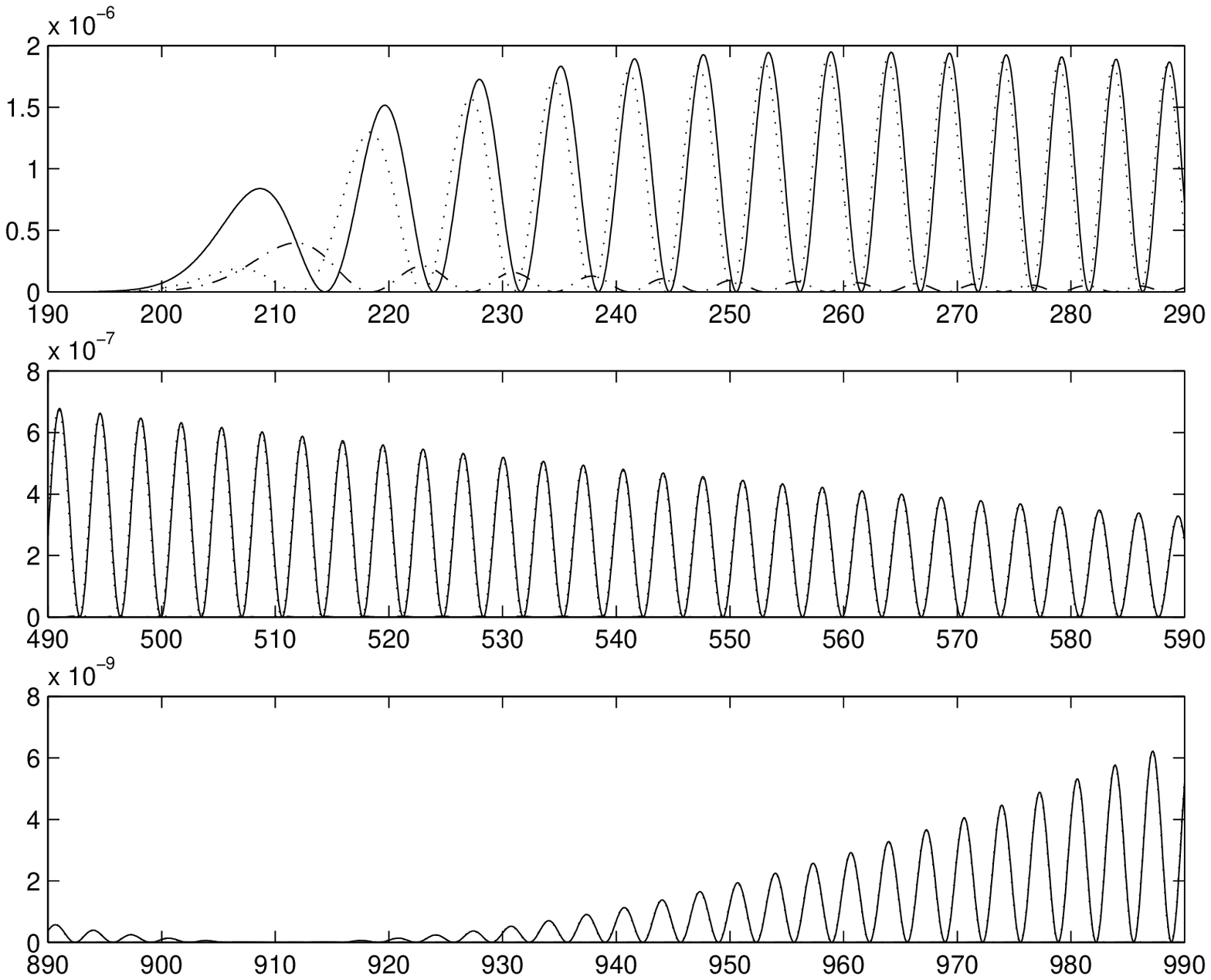}} 
\caption{The same as in figure \ref{integrandc2dp}, for the multipole 
$\ell =200$. The main contribution to this 
multipole comes from relatively short waves, where term B dominates. The 
modulating term and the Bessel functions combine 
in such a manner, that the area under the curve at long and intermediate 
wavelengths is much smaller than at short 
wavelengths. The situation is opposite to the case of small multipoles.}  
\label{integrandc200dp} 
\end{figure}

$\ell \geq 2$: There exists a qualitative difference between small multipoles 
$\ell \approx 2-10$ and relatively 
large multipoles $\ell \approx 100-200$, in the sense of the different roles 
played by terms B and D (see the 
decomposition of $D_{\ell \geq 2}\!(n)$ in formula (\ref{eq:differentDl}) ). 
In the case of small multipoles, the 
dominant contribution to the integral comes from long and intermediate 
wavelengths, where the most important term in 
$D_{\ell \geq 2}\!(n)$ is term D. In the case of relatively large 
multipoles, the dominant contribution comes from 
short wavelengths, where the most important term is term B. In figures 
\ref{integrandc2dp}, \ref{integrandc10dp} and 
\ref{integrandc200dp}, we present by a solid line the behaviour of the total 
integrands for $C_{2}$, $C_{10}$ and 
$C_{200}$ at different $n$'s. The values of the parameters are 
$z_{{\mathrm{eq}}}=7000$, 
$z_{{\mathrm{dec}}}=1000$ and $\beta=-2$. At the same 
graphs, we have also plotted with a dashed-dotted line the integrand that 
would result from neglecting term B, and with a 
dotted line the integrand that would result from neglecting term D.  

From figures \ref{integrandc2dp} and \ref{integrandc10dp}, it is evident that 
the dominant contribution to $C_{2}$ and 
$C_{10}$ comes from wavelengths longer than and comparable to the Hubble 
radius. In this regime, the role of term B is 
unimportant. Although term B will eventually dominate in the short wave 
regime, the contribution of that regime is 
negligible. At the same time, the modulating function in the most sensitive 
part of integration is close to 1. Therefore, 
the inaccurate formula (\ref{eq:mutilatedCl}) gives roughly correct numerical 
values for small multipoles. On the other 
hand, as figure 
\ref{integrandc200dp} shows, the main contribution to the relatively large 
multipoles comes from short scales, where 
term B is always dominant. The neglect of term B gives wrong results.

Summarising, one can say the following. The major contribution to the 
monopole moment comes from wavelengths larger than 
and comparable to the Hubble radius. The actually measured isotropic
temperature $T_0$ contains this small contribution. The major contribution to 
the dipole moment comes from wavelengths 
significantly shorter than the 
Hubble radius, but still far away from the onset of nonlinearities. 
Numerically, 
the dipole moment is much larger than the 
monopole and the $\ell \geq 2$ multipoles, as seen 
from the comparison of figures \ref{integrandc0} and 
\ref{integrandc2dp}--\ref{integrandc200dp} with \ref{integrandc1}. To 
illustrate the greater sensitivity of the dipole to the short waves, one can 
consider the specific example of $\beta =-2$, 
$z_{{\mathrm{eq}}}=7000$ and $z_{{\mathrm{dec}}}=1000$. The ratio of the 
dipole to the quadrupole 
is $C_{1}/C_{2} \sim 1.4\cdot 10^{5}$. This 
ratio in the absence of the modulating function is $\sim 10^{6}$. The major 
contribution to the $\ell \in [2,20]$ 
multipoles comes from wavelengths longer than and comparable to the Hubble 
radius. The major contribution to the 
relatively large multipoles comes from short wavelengths, where term B 
dominates. Still, these multipoles are several 
orders of magnitude smaller than the dipole.

\subsection{Variation of the parameters} 
\label{subsec-FIFTHA}
\begin{figure}
\centerline{\epsfxsize=12cm \epsfbox{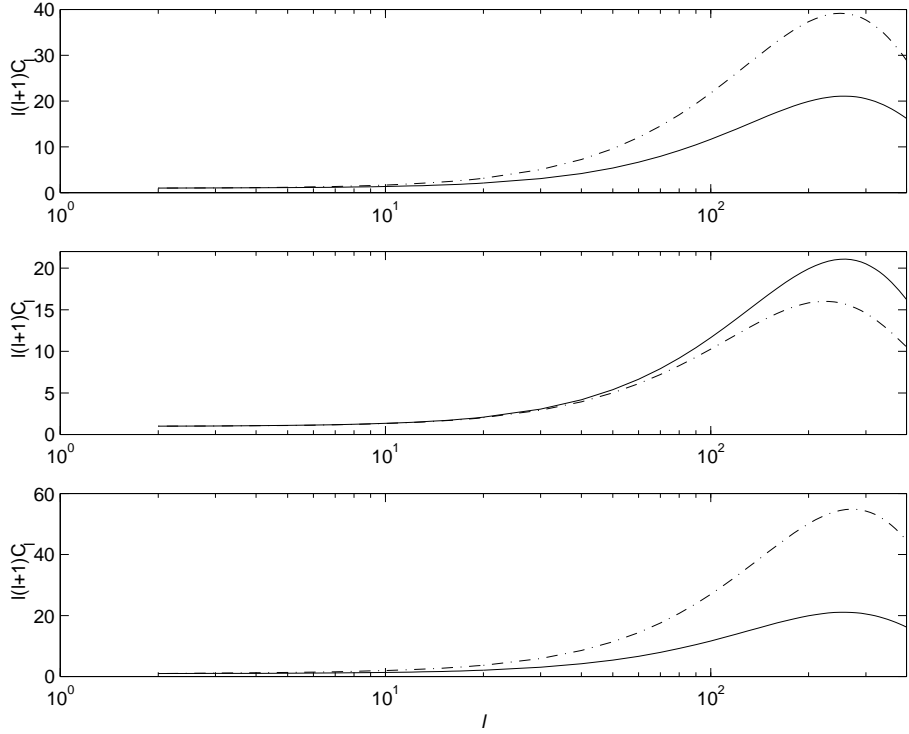}} 
\caption{The effect of change of different parameters. The quantity 
$\ell(\ell+1) C_{\ell}$ is normalised to 1 at 
$\ell=2$. In all graphs, the solid line corresponds to 
$z_{{\mathrm{dec}}}=1000$, 
$z_{{\mathrm{eq}}}=7000$, $\beta=-2$. i) Top graph: the 
effect of decreasing $z_{{\mathrm{dec}}}$. The dashed-dotted line corresponds 
to 
$z_{{\mathrm{dec}}}=500$, $z_{{\mathrm{eq}}}=7000$, $\beta=-2$. The peak 
is displaced to the left and its height increases. ii) Middle graph: the 
effect of decreasing $z_{{\mathrm{eq}}}$. The dashed-dotted 
line corresponds to $z_{{\mathrm{dec}}}=1000$, $z_{{\mathrm{eq}}}=5000$, 
$\beta=-2$. The peak is 
displaced to the left and its height 
decreases. iii) Bottom graph: the effect of increasing $\beta$. The 
dashed-dotted line corresponds to $z_{{\mathrm{dec}}}=1000$, 
$z_{{\mathrm{eq}}}=7000$, $\beta=-1.9$. The peak is displaced to the right 
and its height increases.} 
\label{pareffectdp} 
\end{figure}

We have demonstrated which wavelength regimes are responsible for the main 
contribution to different multipoles. While 
calculating the integral (\ref{eq:Clfinal}), one particular (typical) choice 
of the parameters has been used: 
$z_{{\mathrm{dec}}}=1000$, $z_{{\mathrm{eq}}}=7000$ and $\beta=-2$. One needs 
to explore how the 
change of these parameters affects the CMBR 
anisotropy pattern. The factor $D_{\ell}(n)$ depends on $z_{{\mathrm{dec}}}$ 
through 
the presence of $\xi$. The modulating 
function depends on $z_{{\mathrm{eq}}}$ through the presence of $\xi_{2}$. The 
primordial spectrum depends on $\beta$. We shall 
separately discuss the effect of the variation of $z_{{\mathrm{dec}}}$, 
$z_{{\mathrm{eq}}}$ and 
$\beta$. The results are presented in figure 
\ref{pareffectdp}. 
 
{\textbf{The effect of}} ${\mathbf{\mathbf{z_{{\mathrm{dec}}}}}}$: The values 
of the 
fixed parameters are $z_{{\mathrm{eq}}}=7000$ and 
$\beta=-2$, while we 
decrease $z_{{\mathrm{dec}}}$ from 1000 (solid line) 
to 500 (dashed-dotted line). The graph shows that the peak is displaced 
towards lower multipoles $\ell$, while the height 
of the peak increases. As already stated, the decrease of 
$z_{{\mathrm{dec}}}$ decreases 
$\xi$. This leads to the increase of term 
B, and explains the rise of the height of the peak. The dipole moment is 
insensitive to the change of $z_{{\mathrm{dec}}}$, since it 
is dominated by term A which is not affected by $z_{{\mathrm{dec}}}$. 
 
{\textbf{The effect of}} ${\mathbf{\mathbf{z_{{\mathrm{eq}}}}}}$: The values 
of the 
fixed parameters are $z_{{\mathrm{dec}}}=1000$ and 
$\beta=-2$, while we 
allow $z_{{\mathrm{eq}}}$ to be 5000 (dashed-dotted line). The graph shows 
that the peak 
is displaced towards lower multipoles 
$\ell$, while the height of the peak decreases. The decrease of 
$z_{{\mathrm{eq}}}$ 
increases $\xi_{2}$, making the modulating 
function decline faster (figure \ref{transdp}). The $C_{\ell}$ will acquire 
smaller values, and therefore the values of 
$\ell(\ell +1)C_{\ell}$ will be smaller. This accounts for the decrease of 
the height of the peak. At the same time, the 
decrease of $z_{{\mathrm{eq}}}$ displaces the first minimum of the modulating 
function 
(and of the modulated spectrum) to smaller 
$n$'s (figure \ref{transdp}). The first maximum of the modulated power 
spectrum will then occur at a lower value of $n$, 
which translates in the displacement of the maximum of 
$\ell(\ell +1)C_{\ell}$ towards smaller $\ell$'s. This is the
dominant reason for the first peak displacement, but we do not exclude
that some additional effects can counterbalance this tendency. 
The decrease of 
$z_{{\mathrm{eq}}}$ decreases the value of the dipole. The ratio 
$C_{1}|_{z_{{\mathrm{eq}}}=7000}\ /\ C_{1}|_{z_{{\mathrm{eq}}}=5000} = 1.25$. 

{\textbf{The effect of}} ${\mathbf{\beta}}$: The values of the fixed 
parameters are $z_{{\mathrm{eq}}}=7000$ and $z_{{\mathrm{dec}}}=1000$, 
while 
we allow $\beta$ to be $-1.9$ (dashed-dotted line). The graph shows that the 
peak is displaced towards higher multipoles 
$\ell$, while the height of the peak increases. The displacement of the peak 
is explained as follows. As shown in figure 
\ref{maximumdp}, the increase of $\beta$ displaces the position of the first 
maximum of the modulated spectrum towards 
larger $n$. As already discussed, the change of $\beta$ does not affect the 
position of the minima. Therefore, the 
maximum of the function $\ell(\ell+1)C_{\ell}$ versus $\ell$ is displaced 
towards higher $\ell$. As for the increase of 
the height of the peak, it is explained as follows. The primordial flat 
spectrum $n^{2\beta +5}$ (top left graph of 
figure \ref{sakharov}, dashed-dotted line) is a smooth line, growing with 
$n$. For a larger $\beta$, the primordial 
spectrum increases faster with $n$, and its plot versus $n$ is a concave 
curve. Therefore, the amplitudes of the 
modulated spectrum are larger for a larger $\beta$, which increases the 
values of the $C_{\ell}$ and the height of the 
peak. The increase of $\beta$ increases also the value of the dipole. The 
ratio 
$C_{1}|_{\beta=-1.9}\ /\ C_{1}|_{\beta=-2} \approx 3$.

\subsection{The inclusion of intrinsic anisotropies} 
\label{subsec-FIFTHB} 

The derivation of the starting formula (\ref{eq:SWformula}) is based on 
solving the differential equation for the 
photons' geodesics \cite{SaWo}. As usual, when defining a solution, one 
should indicate the initial conditions. In our 
problem, this translates into the intrinsic CMBR anisotropies. These 
anisotropies reflect the temperature inhomogeneities already 
present on the last scattering surface. There are many elaborate treatments 
dealing with this subject (for example, Doroshkevich \& Schneider 
\shortcite{DoSch}, Hu \& Sugiyama \shortcite{HuSu}). We shall use a simplified 
model (see, 
e.g. Padmanabhan \shortcite{Padma}), where the intrinsic 
anisotropy is related to the radiation density contrast by the simple 
relationship  
\[ \left( \frac{\delta T}{T}\right)_{{\mathrm{in}}} = \frac{1}{4} 
\left.\left(\frac{\delta \rho}{\rho}\right)_{{\mathrm{rad}}}\right|_{\eta = 
\eta_{{\mathrm{E}}}}.\] 
Deeply in the radiation-dominated era, photons are tightly coupled to baryons 
through Thomson scattering. It can be seen 
that the radiation density contrast can be related to the baryon density 
contrast by  
\[ \left(\frac{\delta \rho}{\rho}\right)_{{\mathrm{rad}}} =  \frac{4}{3} 
\left(\frac{\delta \rho}{\rho}\right)_{{\mathrm{b}}}.\] 
Assuming that the dominant matter component governing the scale factor at the 
matter-dominated stage is a sort of dark 
matter, the baryon density contrast can be related to the dark matter density 
contrast by 
\[ \left(\frac{\delta \rho}{\rho}\right)_{{\mathrm{b}}} = 
\frac{z_{{\mathrm{dec}}}}{z_{{\mathrm{eq}}}}\ 
\left(\frac{\delta \rho}{\rho}\right)_{{\mathrm{dm}}}.\] 
Therefore, one can relate the intrinsic anisotropies to the total density 
variation at the last scattering surface: 
\[ \left( \frac{\delta T}{T}\right)_{{\mathrm{in}}} = \frac{1}{3} 
\frac{z_{{\mathrm{dec}}}}{z_{{\mathrm{eq}}}}\ 
\left. \left(\frac{\delta \rho}{\rho}\right) \right|_{\eta = 
\eta_{{\mathrm{E}}}}.\]  
This intrinsic part should be added to the gravitational part of the previous 
Sections, and this total temperature 
variation should be decomposed over spherical harmonics. One then repeats the 
procedure already performed in the absence 
of the intrinsic anisotropy, to derive the full formula (containing both 
gravitational and intrinsic parts) for the  
multipole moments $C_{\ell}$. This is again equation (\ref{eq:Clfinal}), 
where the expression (\ref{eq:Dl}) for 
$D_{\ell}(n)$ is modified to  
      \begin{equation} 
         D_{\ell}(n) = \frac{1}{3}\ n\ \delta_{\ell 1} +  n(1- \xi)\ 
\frac{1}{2\ell +1} 
[ (\ell +1)\ j_{\ell +1}\!(n\xi) - \ell\ j_{\ell -1}\!(n\xi)] + 
\delta_{\ell 0}  - 
\left[ 1 + \frac{3}{10}\ \frac{1}{6}\ (n\xi_{2})^{2} \right]\ j_{\ell}(n\xi).
   \label{eq:totalDl} 
      \end{equation} 
The inclusion of the intrinsic part causes the further increase of the 
multipole moments $C_{\ell}$, and the increase of 
the height of the first peak in the CMBR distribution. If plasma fluid
at the last scattering surface retains appreciable velocity with
respect to chosen by us coordinate system, which is synchronous and
comoving with gravitationally dominant matter, this will give rise to
further corrections. As will be argued in 
the next Section, the observed first peak 
seems already to be too high to be explained, together with other multipoles, 
by density perturbations alone. 

\section{A first comparison with observations} 
\label{sec-SIXTH}

The current CMBR observational results can be summarised as follows.\\
 
1) The COBE observations \cite{Benn} have shown the presence of a 
`plateau' in the multipole region 
$\ell \in [2,20]$. As a characteristic value in this region, we shall use the 
quantity 
$\ell(\ell+1){\mathbf{C}}_{\ell}|_{\ell =10}$. By the bold-face symbol 
${\mathbf{C}}_{\ell}$ we denote the value of the 
multipole moment $\ell$ that follows from observations. (The lack of
ergodicity on a 2-sphere \cite{LPGMar} prevents one from precise extraction
of statistical $C_{\ell}$ from a single, even if arbitrarily accurate, 
map of the sky, 
but this is not a matter of concern for us here.) The COBE result shows that 
\[ \sqrt{\frac{\ell(\ell+1){\mathbf{C}}_{\ell}|_{\ell =10}}{2\pi}} 
\cdot T_{0} = 27.7^{+3.9}_{-4.5}\ {\mathrm{\mu K}},\] 
where $T_{0}$ is the background, isotropic, temperature of the CMBR. 
\footnote{In many 
topics related to CMBR experiments and 
observational data, we have resorted to the web-site 
http://www.hep.upenn.edu/~max/cmb/experiments.html.} This gives the value of 
$\ell(\ell+1){\mathbf{C}}_{\ell}|_{\ell =10}$:  
      \begin{equation} 
         \ell(\ell+1){\mathbf{C}}_{\ell}|_{\ell =10} \approx 6.4 
\cdot 10^{-10} .        \label{eq:obs10} 
      \end{equation} 
2) The dipole anisotropy was also measured by COBE. According 
to Bennett et al. \shortcite{Benn}, 
\[ \sqrt{\frac{(2\ell  +1){\mathbf{C}}_{\ell}|_{\ell =1}}{4\pi}} \cdot T_{0} 
= 3.353 \pm 0.024\ {\mathrm{mK}} ,\] 
leading to 
      \begin{equation} 
{\mathbf{C}}_{1} \approx 6.3 \cdot 10^{-6} .        \label{eq:obs1} 
      \end{equation} 
3) All the CMBR observations performed at the region of higher $\ell$'s have 
suggested the presence of a peak. The 
most recent and accurate experiments BOOMERanG \cite{deBer} and MAXIMA-1 
\cite{Hana} gave the location and height of 
this peak with some precision. According to these results, the peak is 
located at $200 \leq \ell _{{\mathrm{peak}}} \leq 220$, with 
$\sqrt{\ell(\ell+1){\mathbf{C}}_{\ell}|_{{\mathrm{peak}}}/(2\pi)} \cdot T_{0}$ 
having 
the value $69 \pm 8 \ {\mathrm{\mu K}}$ 
\cite{deBer} or $78 \pm 6 \ {\mathrm{\mu K}}$ \cite{Hana}. 
For the sake of this discussion, we adopted the values of Hanany et al. 
\shortcite{Hana},  
      \begin{equation} 
          \ell_{{\mathrm{peak}}} \simeq 220   \label{eq:peakposition} 
      \end{equation}  
and  
\[ \sqrt{\frac{\ell(\ell+1){\mathbf{C}}_{\ell}|_{{\mathrm{peak}}}}{2\pi}} 
\cdot T_{0} = 
78 \pm 6 \ {\mathrm{\mu K}}.\] 
Therefore, we shall use as a characteristic value in this region the quantity 
      \begin{equation} 
\ell(\ell+1){\mathbf{C}}_{\ell}|_{{\mathrm{peak}}} \approx 5.1 \cdot 10^{-9} . 
       \label{eq:obspeak} 
      \end{equation} 
 
We accept the position of the peak (\ref{eq:peakposition}), and 
we regard the three numbers (\ref{eq:obspeak}), 
(\ref{eq:obs1}), (\ref{eq:obs10}) as three observational points to 
which we try to fit (admittedly, in a somewhat simplistic manner) the 
theoretical curves. The curves for 
$\ell(\ell +1)C_{\ell}$ are constructed with the help of formulae 
(\ref{eq:Clfinal}) and (\ref{eq:totalDl}), where the 
upper limit of integration 
is taken at $n_{{\mathrm{max}}} = 1260$. The strategy of the fitting is as 
follows.   

\begin{figure}
\centerline{\epsfxsize=15cm \epsfbox{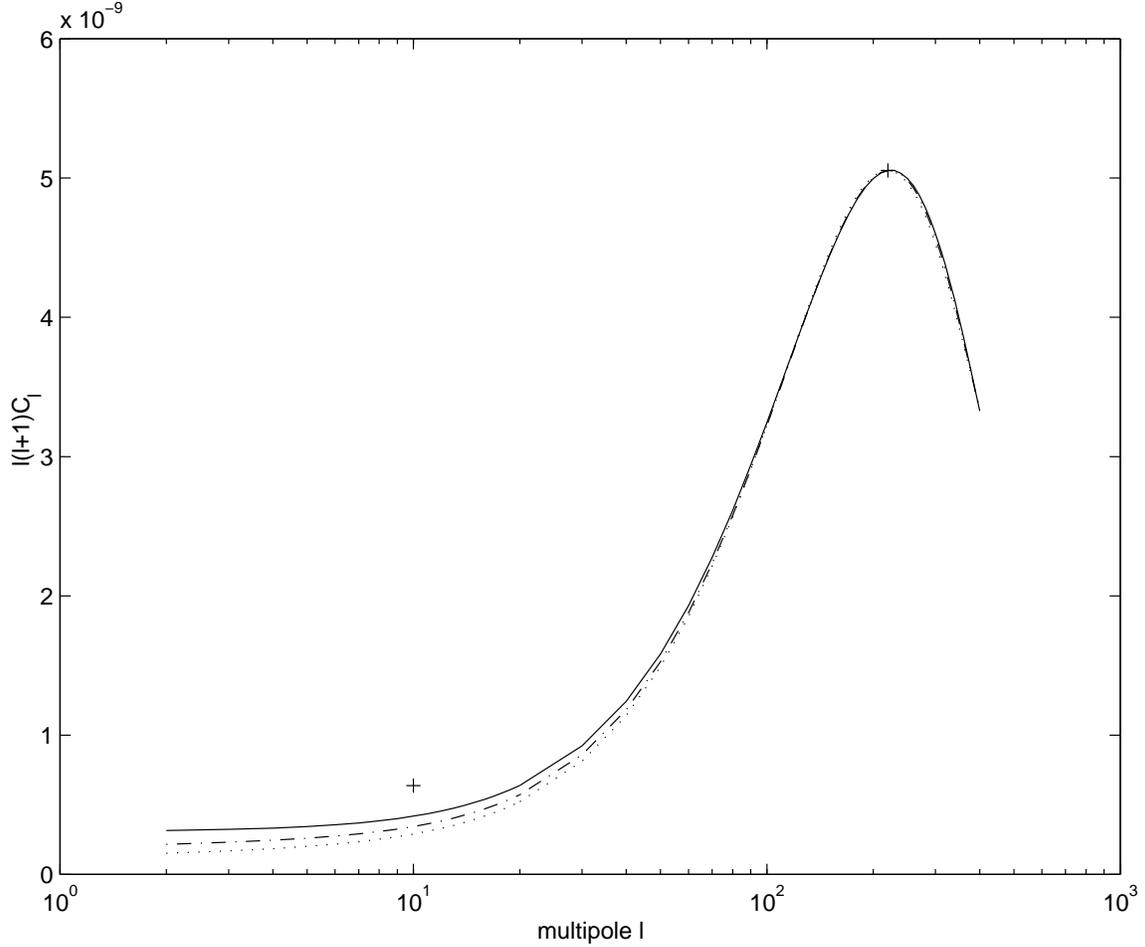}}
\caption{Comparison of theoretical models, based on density perturbations 
only, with observations. The solid line 
corresponds to the set of parameters $(5000,1000,-2)$, the dashed-dotted to 
$(4500,1000,-1.95)$ and the dotted to 
$(4000,1000,-1.9)$. The crosses show the observed values of 
$\ell(\ell+1){\mathbf{C}}|_{\ell=10}$ and 
$\ell(\ell+1){\mathbf{C}}_{\ell}|_{{\mathrm{peak}}}$. By fitting the 
theoretical peak to 
the observed peak, we notice that there 
is a deficit at the plateau.}
\label{compnogw}
\end{figure}

There are several combinations of the parameters 
$(z_{{\mathrm{eq}}},\ z_{{\mathrm{dec}}},\ \beta)$ 
(some of them are listed in table 
\ref{comparisondp}) for 
which the position of the peak is at (\ref{eq:peakposition}) and the height 
of the peak is at (\ref{eq:obspeak}). 
At the same time, we check that the calculated statistical dipole is 
close (deviation less than 5 per cent) to the observed 
value (\ref{eq:obs1}). A drastic adjustment of the calculated 
dipole with the help of a `peculiar' velocity of 
the individual observer, that is, allowing the observer to move 
with a large velocity with respect to the matter-dominated fluid 
(i.e. with respect to the comoving coordinate system) seems to be 
artificial. As the statistical
dipole is already quite large, it seems especially
unnatural to try to do this.  As a guideline in the choice of 
$z_{{\mathrm{dec}}}$ and 
$z_{{\mathrm{eq}}}$, we use the fact that 
$z_{{\mathrm{dec}}}$ should be close to $1000$ and $z_{{\mathrm{eq}}}$ should 
be somewhere near to 
$10^{4}$. The decoupling red-shift $z_{{\mathrm{dec}}}$ 
certainly depends on the baryon content
of the Universe $\Omega_{b}$, but it can also depend on a complicated 
ionisation
history of the decoupling era. The equality red-shift $z_{{\mathrm{eq}}}$ 
certainly
depends on the total matter content of the Universe (baryons, cold dark
matter, etc.) but it also depends on its radiation content 
(CMB photons, possible contribution of massless neutrinos, etc.). 
In addition, the total density parameter $\Omega$ can also deviate from the 
postulated
here $\Omega = 1$. This may change $z_{{\mathrm{eq}}}$ without seriously 
affecting our
solutions for the gravitational field perturbations $h_{ij}$ and subsequent
formulae, derived, strictly speaking, for the case $\Omega = 1$.  
It would be a difficult problem to try to decipher all the relevant 
quantities
from just the two numbers $z_{{\mathrm{dec}}}$ and $z_{{\mathrm{eq}}}$. 
However, we will operate 
with
these parameters, since this is what the calculations immediately require 
us
to specify. As for $\beta$, we have considered the 
three different cases $\beta =$ $-2$, $-1.95$, $-1.9$. Since $K\!(\beta)$ is 
not very sensitive to $\beta$ in this 
interval, we 
have used one and the same formula (\ref{eq:K}) for all the three cases. 
Then, every fit of the peak and the dipole to 
their observed values defines its own value for the ratio 
$\Lambda \equiv (l_{{\mathrm{Pl}}}/l_{0})^{2}$ and for the calculated 
quantity $\ell(\ell+1)C_{\ell}|_{\ell=10}$. This quantity is finally compared 
with the observed value (\ref{eq:obs10}). 
The result of this comparison shows that the theoretical 
$\ell(\ell+1)C_{\ell}|_{\ell=10}$ is systematically smaller than 
the observed $\ell(\ell+1){\mathbf{C}}|_{\ell=10}$, for all considered 
combinations of the parameters 
$(z_{{\mathrm{eq}}},\ z_{{\mathrm{dec}}},\ \beta)$. The observational points 
should be 
surrounded by their error bars, but in order not to proliferate the already
existing uncertainties, we focus on the mean values. One can say that, 
despite of the still allowed variation of the 
parameters, the calculated 
`plateau' is persistently lower than it should be. These calculations are 
visualised in figure \ref{compnogw}. 

We believe that this deficit at small multipoles serves as an indication 
that 
density perturbations alone can hardly be responsible for all of
the CMBR anisotropies. The most natural 
way to proceed is to include gravitational waves and repeat this scheme.

\begin{table}
   \caption{Comparison of some theoretical models, strictly satisfying the 
requirements on the position and the height of 
the first peak, with observations.}
   \label{comparisondp}
   \begin{tabular}{@{}lccccccc}
   $z_{{\mathrm{eq}}}$&$z_{{\mathrm{dec}}}$&$\beta$&$\Lambda$&$\ell(\ell+1)C_{\ell}|_{\ell=10}$&${\mathbf{C}}_{10}/C_{10}$&$C_{1}$&${\mathbf{C}}_{1}/C_{1}$ \\  \hline
   5000&1000&-2.00&$7.7\cdot 10^{-9}$&$4.2\cdot 10^{-10}$&1.5&$6.1\cdot 10^{-6}$&1.04 \\  
   4500&1000&-1.95&$5.0\cdot 10^{-9}$&$3.4\cdot 10^{-10}$&1.9&$6.1\cdot 10^{-6}$&1.03 \\  
   4000&1000&-1.90&$3.3\cdot 10^{-9}$&$2.9\cdot 10^{-10}$&2.2&$6.1\cdot 10^{-6}$&1.03 \\  \hline
   \end{tabular}

\end{table}

\section{The gravitational wave contribution to the CMBR anisotropies}
\label{sec-SEVENTH}

Gravitational waves are inevitably generated by the strong variable 
gravitational field of the early Universe (see, for 
example, Grishchuk et al. \shortcite{LPGetal} and references therein). The 
preliminary calculations 
show \cite{LPG2} that the contribution 
of quantum-mechanically generated gravitational waves to the small multipoles 
should be of the same order of magnitude 
and, numerically, somewhat larger than the contribution of density 
perturbations generated by the same mechanism. In this 
paper, however, we follow a phenomenological approach and do not not focus on 
the origin of cosmological perturbations 
and consequences of the particular generating mechanism. We introduce, 
instead, a phenomenological parameter $\varpi$, 
which regulates the relative contributions of gravitational waves and density 
perturbations. In effect, we want to find 
the value of $\varpi$ which allows to raise the plateau in figure 
\ref{compnogw} to the observationally required level, 
at the expense of gravitational waves. At the same time, the fits to the peak 
and the dipole are required to be as good 
as before. It is already clear from table \ref{comparisondp} and  figure 
\ref{compnogw} that the required amount of 
gravitational waves, as compared with density perturbations, is within a 
factor of 2 at $\ell =10$. This seems to be in 
agreement with theoretical expectations, but we shall go into a more detailed 
analysis in Section \ref{sec-EIGHTH}. 

In this paragraph, we shall only remind the fundamentals about primordial 
gravitational waves. Although we follow the 
quantum mechanical derivation, the final formula (\ref{eq:Clgw}) for 
$C_{\ell}$ is true (up to the overall numerical 
coefficient) for a broad class of gravitational waves. The Heisenberg operator 
for the gravitational wave metric 
perturbations is 
     \begin{equation}
         h_{ij}(\eta,{\mathbf{x}}) = l_{{\mathrm{Pl}}}
\frac{\sqrt{16\pi}}{(2\pi)^{3/2}}\ 
\int_{-\infty}^{\infty} {\mathrm{d}}^{3}{\mathbf{n}}\ \frac{1}{\sqrt{2n}}\ 
\sum_{s=1}^{2}\ \stackrel{s}{p}_{ij}\!\!({\mathbf{n}})\ 
\left[\stackrel{s}{h}_{n}\!\!(\eta)\ \stackrel{s}{c}_{{\mathbf{n}}}\!\!(0)\ 
{\mathrm{e}}^{{\mathrm{i}}{\mathbf{n\cdot x}}} + 
\stackrel{s}{h}_{n}\!\!\!\!\!^{\ast}(\eta)\ 
\stackrel{s}{c}_{{\mathbf{n}}}\!\!\!\!\!^{\dagger}(0)\ 
{\mathrm{e}}^{-{\mathrm{i}}{\mathbf{n\cdot x}}} \right],  \label{eq:quantumgw}
     \end{equation}
where, again, $l_{{\mathrm{Pl}}}$ comes from quantum normalisation, and the 
transverse traceless polarisation tensors 
$\stackrel{s}{p}_{ij}\!\!({\mathbf{n}})$ 
satisfy the relations 
\[ \stackrel{s}{p}_{ij}\!\!({\mathbf{n}})\ 
\stackrel{s'}{p}\!\!^{ij}\!({\mathbf{n}})=2\delta_{ss'},\ 
\stackrel{s}{p}_{ij}\!\!({\mathbf{n}})=\stackrel{s}{p}_{ij}\!\!({\mathbf{-n}}),
\ \stackrel{s}{p}_{ij}n^{j}=0,\ \stackrel{s}{p}_{ij}\delta^{ij}=0. \] 
The time evolution of each of the amplitudes $h_{n}\!(\eta)$ for $s=1,\ 2$ is 
given by one and the same equation with the 
same 
initial conditions, so one deals with one and the same $h_{n}\!(\eta)$. After 
solving for $h_{n}\!(\eta)$ at the 
three stages (\ref{eq:scalefactorsfinal}) and continuously joining the 
solutions, one arrives at a `growing' and a 
`decaying' solution in the matter-dominated era. The growing solution is 
given by the expression 
     \begin{equation}
         h_{n}\!(\eta) \approx {\mathrm{i}}\ {\mathrm{e}}^{{\mathrm{i}} 
n\eta _{0}}\ 3\Psi\!(\beta)
\ (-1)^{-\beta}\ l_{0}^{-1}\ n^{\beta +1}\ \frac{j_{1}\!(n(\eta -\eta_{m}))}
{n(\eta -\eta_{m})} ,  \label{eq:metricfourriergw}
     \end{equation}
where the quantity $|\Psi\!(\beta)|^2$ is unity for $\beta = -2$, and we 
neglect its weak $\beta$-dependence for other 
values of $\beta$ considered here. For a reference, $|\Psi\!(\beta)|^2$ is 
related with $K\!(\beta)$ discussed in Section 
\ref{sec-THIRD} by  
\[ K\!(\beta) = 2\pi |\Psi\!(\beta)|^{2} \left(\frac{l_{{\mathrm{Pl}}}}{l_{0}}\right)^{2}.
 \]
Combining (\ref{eq:metricfourriergw}) with (\ref{eq:quantumgw}), using the 
Sachs-Wolfe formula (\ref{eq:SWformula}), and 
repeating the same steps that have led to the formula for $C_{\ell}$'s caused 
by density perturbations, one arrives at 
\cite{LPG1} 
   \begin{equation}
     C_{\ell} \approx 4\cdot 9|\Psi\!(\beta)|^{2}\ 
\left( \frac{l_{{\mathrm{Pl}}}}{l_{0}}\right)^{2}\ (\ell -1) \ell (\ell +1) 
(\ell +2) 
\int_{0}^{+\infty } n^{2(\beta +2)}\ f_{\ell}(n)\ \frac{{\mathrm{d}}n}{n} ,    
\label{eq:Clgw}
   \end{equation}
where
         \[  f_{\ell}(n) \equiv \left| \int_{0}^{n\xi} 
\frac{j_{\ell}\!(x)}{x^{2}}\ \frac{j_{2}\!(n-x)}{(n-x)} 
{\mathrm{d}}x \right|^{2}  .\] 
As (\ref{eq:Clgw}) shows, gravitational waves contribute to multipoles 
$\ell \geq 2$, in contrast to primordial density 
perturbations, which contribute to all multipoles. 

\begin{figure}
\centerline{\epsfxsize=12cm \epsfbox{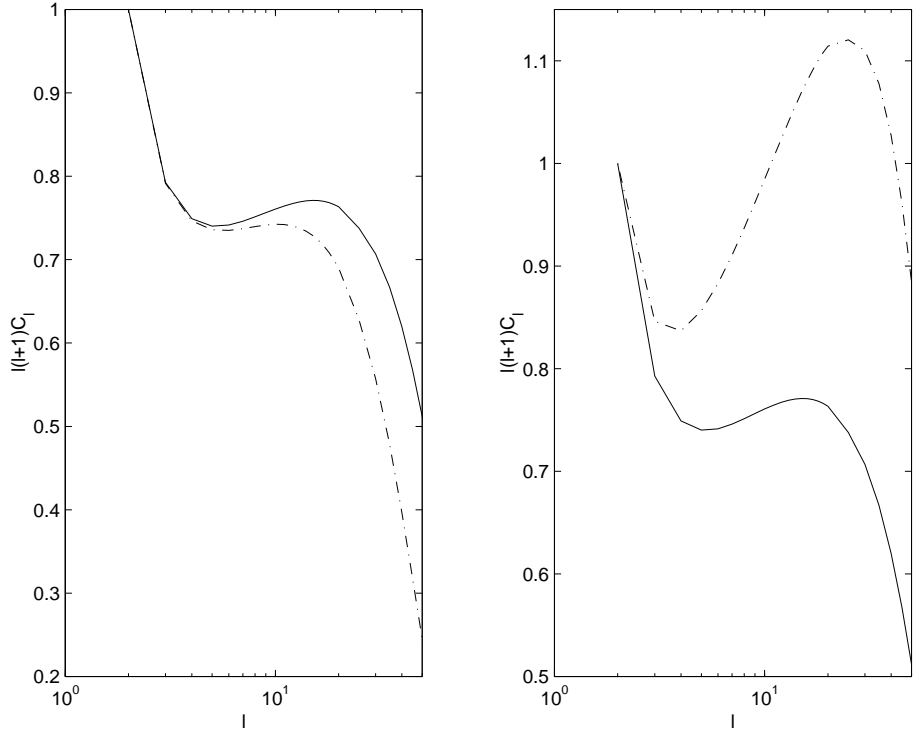}}
\caption{The effect of the variation of the different parameters on 
$\ell(\ell+1)C_{\ell}$. In both graphs, the solid 
line corresponds to 
$z_{{\mathrm{dec}}}=1000,\ z_{{\mathrm{eq}}}=7000,\ \beta=-2$. i) Left graph: 
the effect of decreasing $z_{{\mathrm{dec}}}$. The 
dashed-dotted line corresponds to 
$z_{{\mathrm{dec}}}=500,\ z_{{\mathrm{eq}}}=7000,\ \beta=-2$. ii) 
Right graph: the effect of increasing 
$\beta$. The dashed-dotted line corresponds to 
$z_{{\mathrm{dec}}}=1000,\ z_{{\mathrm{eq}}}=7000,\ \beta=-1.9$. The local 
maximum is displaced 
to the right and its height increases, exceeding the value of 
$\ell(\ell+1)C_{\ell}|_{\ell=2}$.}
\label{pareffectgw}
\end{figure}

We shall now present the multipole moments caused by gravitational waves. The 
solid line in both graphs of 
figure \ref{pareffectgw}, shows the behaviour of $\ell(\ell+1)C_{\ell}$ 
versus $\ell$ 
for $z_{{\mathrm{eq}}}=7000$, $z_{{\mathrm{dec}}}=1000$ and $\beta =-2$. The 
quantity 
$\ell(\ell+1)C_{\ell}|_{\ell =2}$ is normalised to 1. The 
higher order multipoles are negligible in comparison with the lower order 
ones, unlike the case of density perturbations. 
This is expected 
because of the different behaviour of the short wavelength density 
perturbations and gravitational waves. At the 
matter-dominated stage, short wavelength density perturbations increase with 
time as $\eta^{2}$, while short wavelength 
gravitational waves decrease with time as $\eta^{-2}$. Figure 
\ref{pareffectgw} also shows the effects of the variation 
of different parameters. As expected, in all cases, gravitational waves 
mostly contribute to the small multipoles. 

\section{A second comparison with observations - Conclusions}
\label{sec-EIGHTH}

We denote by $C^{{\mathrm{dp}}}_{\ell}$ and $C^{{\mathrm{gw}}}_{\ell}$ the 
multipole moments 
induced by density perturbations and gravitational 
waves, respectively. $C^{{\mathrm{dp}}}_{\ell}$ is given by 
(\ref{eq:Clfinal}), whereas 
$C^{{\mathrm{gw}}}_{\ell}$ is given by (\ref{eq:Clgw}). 
We introduce a numerical parameter $\varpi$ in such a way that the total 
$\ell(\ell+1)C_{\ell}$ is given by 
       \begin{equation}
         \ell(\ell+1)C_{\ell} = \ell(\ell+1)[C^{{\mathrm{gw}}}_{\ell} + 
\varpi \cdot C^{{\mathrm{dp}}}_{\ell}] . \label{eq:defvarpi}
       \end{equation}
The value of $\varpi$ would be strictly 1, if one could guarantee that the 
idealised model that has led to formulae 
(\ref{eq:Clfinal}), (\ref{eq:Clgw}) has accurately taken into account all 
physical processes that might have affected 
density perturbations and gravitational waves throughout their entire 
evolution, from the moment they existed in a vacuum 
quantum state and up to the era of decoupling of CMB radiation from matter. 
Since one may have legitimate doubts in such 
accuracy, 
the correction parameter $\varpi$ may be thought of as a measure of a 
necessary modification (but only in the form of an 
overall numerical 
coefficient) to our theoretical framework. 

\begin{figure}
\centerline{\epsfxsize=15cm \epsfbox{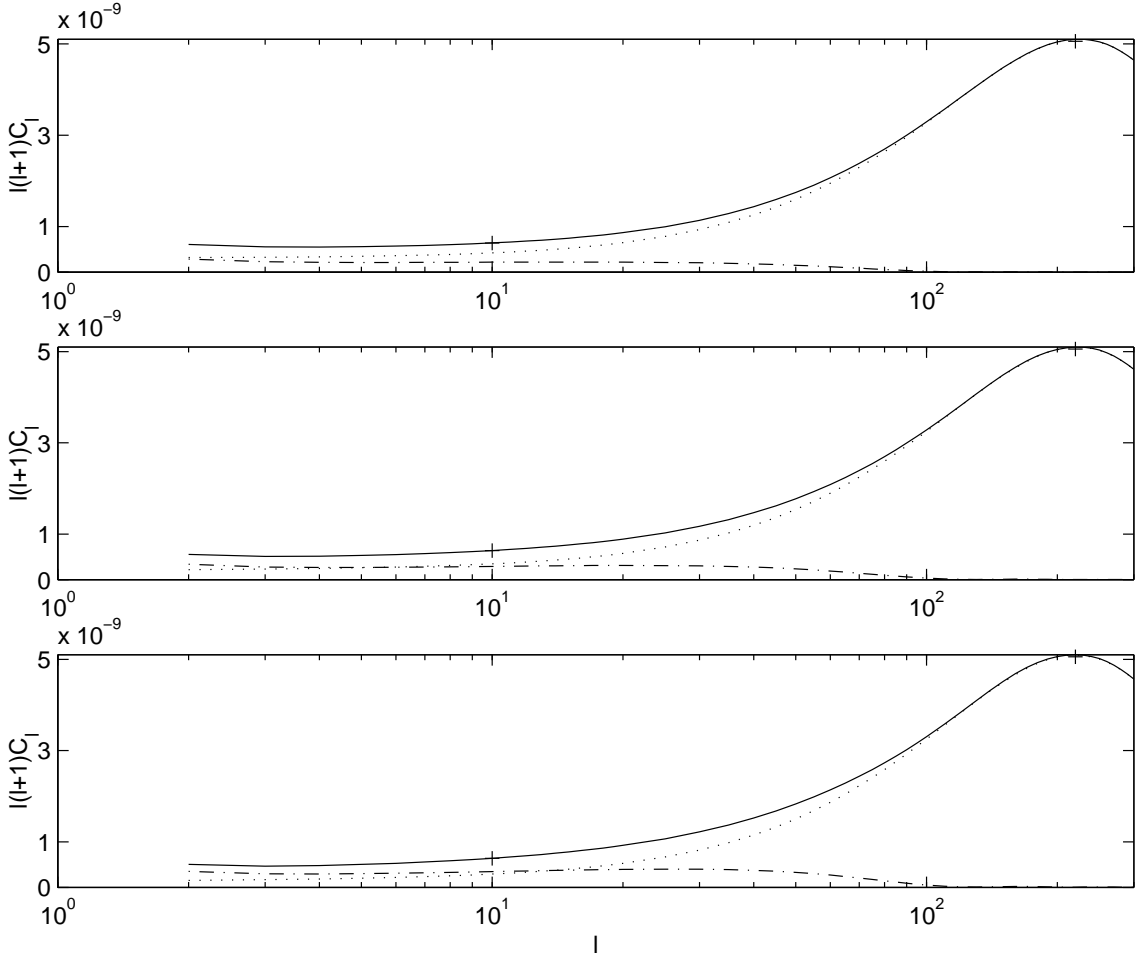}}
\caption{Comparison of the theoretical model, including density perturbations 
and gravitational waves, with observations. 
The dotted line shows the contribution of the corrected density perturbations 
alone, the dashed-dotted line shows the 
contribution of gravitational waves alone, and the solid line is the total 
$\ell(\ell+1)C_{\ell}$. The top graph 
corresponds to the set of parameters $(5000,1000,-2)$, the middle graph to 
$(4500,1000,-1.95)$, and the bottom graph to 
$(4000,1000,-1.9)$. The crosses, again, correspond to the observed values of 
$\ell(\ell+1){\mathbf{C}}|_{\ell=10}$ and 
$\ell(\ell+1){\mathbf{C}}|_{{\mathrm{peak}}}$. In all cases, the contribution 
of 
gravitational waves to the plateau is comparable to 
that of density perturbations, and for larger values of $\beta$ the 
contribution of gravitational waves is numerically 
greater than that of density perturbations.}
\label{compinclgw}
\end{figure}

As was shown in Section \ref{sec-SEVENTH}, gravitational waves do not 
contribute to the dipole, and do not practically 
contribute to the peak. Therefore, we require the quantity 
$\varpi \cdot \ell(\ell+1)C^{{\mathrm{dp}}}_{\ell}|_{{\mathrm{peak}}}$ to be 
equal 
to the characteristic value (\ref{eq:obspeak}). At the same time, we require 
$\ell(\ell+1)C_{\ell}|_{\ell=10}$, defined 
by setting $\ell =10$ in (\ref{eq:defvarpi}), to be equal to the 
characteristic value (\ref{eq:obs10}). This comparison 
revises the ratio $\Lambda= (l_{{\mathrm{Pl}}}/l_{0})^2$ and determines the 
parameter 
$\varpi$. We then construct the ratio 
$\ell(\ell+1)C^{{\mathrm{gw}}}_{\ell}|_{\ell=10}/(\varpi \cdot \ell(\ell+1)C_{\ell}^{{\mathrm{dp}}}|_{\ell=10})= C^{{\mathrm{gw}}}_{10}/(\varpi \cdot C^{{\mathrm{dp}}}_{10})$, 
which shows the relative contributions of gravitational waves and density 
perturbations to the small multipoles. 
Finally, we compare the value $\varpi \cdot C_{1}$ with the observed value 
(\ref{eq:obs1}) of the dipole. The results of 
this analysis are shown in table \ref{fullcomparison}, while a visual 
representation is given in figure \ref{compinclgw}.

\begin{table}
   \caption{Comparison of the theoretical model with observations, for the 
sets of parameters $(5000,1000,-2)$, 
$(4500,1000,-1.95)$ and $(4000,1000,-1.9)$. Primordial gravitational 
waves have now been taken into account. The position and the height of the 
first peak, as well as the dipole moment and the plateau, are in an excellent 
fit to the observations.}
   \label{fullcomparison}
\begin{small}
   \begin{tabular}{@{}lccccccccc}
   $z_{{\mathrm{eq}}}$&$z_{{\mathrm{dec}}}$&$\beta$&$\Lambda$&$\varpi$&$\varpi \cdot \ell(\ell+1)C^{{\mathrm{dp}}}_{\ell}|_{\ell=10}$&$\ell(\ell+1)C^{{\mathrm{gw}}}_{\ell}|_{\ell=10}$&$C^{{\mathrm{gw}}}_{10}/(\varpi \cdot C^{{\mathrm{dp}}}_{10})$&$\varpi \cdot C^{{\mathrm{dp}}}_{1}$&${\mathbf{C}}_{1}/(\varpi \cdot C^{{\mathrm{dp}}}_{1})$ \\  \hline
   5000&1000&-2.00&$2.6\cdot 10^{-10}$&29.2&$4.1\cdot 10^{-10}$&$2.2\cdot 10^{-10}$&0.5&$6.1\cdot 10^{-6}$&1.04 \\  
   4500&1000&-1.95&$2.9\cdot 10^{-10}$&17.2&$3.4\cdot 10^{-10}$&$2.9\cdot 10^{-10}$&0.8&$6.1\cdot 10^{-6}$&1.03 \\  
   4000&1000&-1.90&$2.8\cdot 10^{-10}$&11.6&$2.9\cdot 10^{-10}$&$3.4\cdot 10^{-10}$&1.2&$6.1\cdot 10^{-6}$&1.03 \\  \hline
   \end{tabular}
\end{small}
  
\end{table}

To summarise, we have revised the formula (\ref{eq:one}) for $C_{\ell}$'s 
caused by density perturbations and derived the 
corrected expression (\ref{eq:Clfinal}). We have shown that the monopole 
moment $C_{0}$ is actually finite and small, 
while the dipole moment is very sensitive to short wavelengths. Because of 
that, the dipole moment is much larger than 
the $\ell =0$ and $\ell \geq 2$ multipoles. 
We have then proceeded to the multipoles $\ell \geq 2$. The correct formula 
predicts growth of the function $\ell(\ell+1) C_{\ell}$ with $\ell$. We have 
illustrated how the presence of the 
modulating function 
$M^{2}\!(n\xi_{2})$ gives rise to the first peak in the CMBR anisotropy 
pattern produced by density perturbations. (It 
would be nice to see further peaks discovered in the positions and with the 
heights suggested by the modulating function, 
but this is the domain of wavelengths where many other physical processes may 
intervene.) A first comparison with 
observations has indicated that the observed multipoles can hardly be 
explained by density perturbations alone, because 
of the deficit in the small $\ell$ plateau. We took a step further and 
included primordial gravitational waves. This 
inclusion gives a picture in better agreement with observations. We explored 
the cases $\beta > -2$ which make the 
primordial spectrum a little bit `blue', thus allowing to avoid the 
infrared divergence of the primordial perturbations 
that occurs for $\beta \leq -2$ (see (\ref{eq:powerB}) and 
(\ref{eq:powerspectrumform})). We conclude that  
models with $\beta > -2$, and therefore with larger contributions of 
gravitational waves, are better than a model 
without gravitational waves and with the flat spectrum $\beta =-2$. This 
analysis is complementary to other studies of 
the CMBR data 
\cite{Albr,Balb1,Balb2,DuNo,Huetal,Kinn,Lang,McGa,MaRiS,PaSe,PeSeH,Pogo,RaSha,ScLaG,TeZa,TeZaH1,TeZaH2}, 
not all of which, though, seem to 
agree with each other.

\section{Acknowledgements}

The authors are grateful to A. Doroshkevich, G. Smoot, L. Piccirillo and 
P. Mauskopf for useful
discussions and correspondence. A. Dimitropoulos was supported by the 
Foundation of State Scholarships of Greece.

\appendix
 
\section{Outline of the calculation of the contribution of term II}
\label{sec-B}

We need to prove that the contribution of term II to  
formula (\ref{eq:Gspherdef}) is given by 
  \begin{equation}    
      {\mathrm{g}}_{\ell m} = \frac{4\pi}{(2\pi)^{3/2}}\ {\mathrm{i}}^{\ell}\ 
\int_{-\infty}^{+\infty} {\mathrm{d}}^{3}{\mathbf{n}}\  B\!_{\mathbf{n}}\ 
n(1-\xi)\ 
\frac{1}{2\ell +1} [ (\ell +1)\ j_{\ell +1}\!(n\xi) -\ell\ j_{\ell -1}\!(n\xi)]
\ Y_{\ell m}^{\ast}\!(\Theta ,\Phi) . \label{eq:B1}
   \end{equation}
The starting point of the calculation is the expression
      \begin{equation}
          G_{II}\!({\mathbf{e}}) = \frac{1}{(2\pi)^{3/2}}\ 
\int _{-\infty}^{+\infty} {\mathrm{d}}^{3}{\mathbf{n}}\ B\!_{\mathbf{n}}\ 
(-{\mathrm{i}})
\ {\mathbf{n\cdot e}}\ (1-\xi)\ 
{\mathrm{e}}^{{\mathrm{i}}{\mathbf{n\cdot e}}\xi}      \label{eq:B2}
      \end{equation}
which needs to be presented as 
   \begin{equation}
      G_{II}\!({\mathbf{e}}) = \sum _{\ell =0}^{\infty}\sum _{m=-\ell}^{\ell} 
{\mathrm{g}}_{\ell m}\ Y_{\ell m}\!(\theta ,\phi) .     \label{eq:B4}
   \end{equation}
Using the representations (\ref{eq:unitvector}) and (\ref{eq:wavevector}) in 
the scalar product ${\mathbf{n\cdot e}}$, the spherical wave expansion 
of a plane wave (\ref{eq:A17}), and the explicit form of 
$Y_{1f}(\theta,\phi)$ (see e.g. page 99 of Jackson \shortcite{Jack}) we derive:
  \begin{equation}
     {\mathrm{i}} {\mathbf{n\cdot e}}\ (1- \xi)\ {\mathrm{e}}^{ {\mathrm{i}} 
{\mathbf{n\cdot e}} \xi} = 
4\pi\ \sum _{\ell =0}^{\infty}\sum _{m=-\ell}^{\ell} \sum_{f=-1}^{1} 
{\mathrm{i}}^{\ell +1}\ \frac{4\pi}{3}\ n(1- \xi)\  
Y^{\ast}_{1 f}(\Theta ,\Phi)
\ Y^{\ast}_{\ell m}\!(\Theta ,\Phi)\ Y_{1 f}(\theta ,\phi)
\ Y_{\ell m}\!(\theta ,\phi).  \label{eq:B3}
  \end{equation}
We now equate the right hand sides of (\ref{eq:B2}) and (\ref{eq:B4}), 
multiply the resulting expression with 
$Y_{\ell ' m'}^{\ast}(\theta ,\phi)$, and use the 
orthogonality condition of the spherical harmonics (e.g. formula (3.55)
in Jackson \shortcite{Jack}). This allows one to write 
   \begin{equation}
      {\mathrm{g}}_{\ell ' m'} = \stackrel{1}{{\mathrm{g}}}_{\ell ' m'} + 
\stackrel{2}{{\mathrm{g}}}_{\ell ' m'} + 
\stackrel{3}{{\mathrm{g}}}_{\ell ' m'}     \label{eq:B5}
   \end{equation}
where
   \begin{eqnarray}
       \stackrel{1}{{\mathrm{g}}}_{\ell ' m'} &=& \frac{4\pi}{(2\pi)^{3/2}}\ 
\sum _{\ell =0}^{\infty} \sum _{m=-\ell}^{\ell} \frac{4\pi}{3}
\ {{\mathrm{i}}} ^{\ell +3}\ \int _{-\infty}^{+\infty} 
{\mathrm{d}}^3{\mathbf{n}}
\ B\!_{\mathbf{n}}\ n(1- \xi)\ j_{\ell}\!(n\xi )\ Y_{1,-1}^{\ast}\!(\Theta ,
\Phi)\ Y_{\ell m}^{\ast}\!(\Theta ,\Phi) \cdot  \nonumber  \\ 
& & \cdot \int _{0}^{\pi} \int _{0}^{2\pi} Y_{1,-1}\!(\theta ,\phi)
\ Y_{\ell m}\!(\theta ,\phi)\ Y_{\ell ' m'}^{\ast}\!(\theta ,\phi)\ 
\sin\!\theta\ {\mathrm{d}}\theta \ {\mathrm{d}}\phi   \label{eq:B6}
   \end{eqnarray}
   \begin{eqnarray}
       \stackrel{2}{{\mathrm{g}}}_{\ell ' m'} &=& \frac{4\pi}{(2\pi)^{3/2}}\ 
\sum _{\ell =0}^{\infty} \sum _{m=-\ell}^{\ell} \frac{4\pi}{3}\ 
{\mathrm{i}} ^{\ell +3}\ \int _{-\infty}^{+\infty} 
{\mathrm{d}}^3 {\mathbf{n}}\ 
B\!_{\mathbf{n}}\ n (1- \xi)\ j_{\ell}\!(n\xi )\ Y_{10}^{\ast}\!(\Theta ,\Phi)
\ Y_{\ell m}^{\ast}\!(\Theta ,\Phi) \cdot  \nonumber  \\
                                            & & \cdot \int _{0}^{\pi} 
\int _{0}^{2\pi} Y_{10}\!(\theta ,\phi)\ Y_{\ell m}\!(\theta ,\phi)\ 
Y_{\ell ' m'}^{\ast}\!(\theta ,\phi)\ \sin\!\theta \ 
{\mathrm{d}}\theta \ {\mathrm{d}}\phi   
\label{eq:B7}
   \end{eqnarray}
\\
\\
   \begin{eqnarray}
       \stackrel{3}{{\mathrm{g}}}_{\ell ' m'} &=& \frac{4\pi}{(2\pi)^{3/2}}\ 
\sum _{\ell =0}^{\infty} \sum _{m=-\ell}^{\ell} \frac{4\pi}{3}\ 
{\mathrm{i}} ^{\ell +3}\ \int _{-\infty}^{+\infty} 
{\mathrm{d}}^3 {\mathbf{n}}\ 
B\!_{\mathbf{n}}\ n(1- \xi)\ j_{\ell}\!(n\xi )\ Y_{11}^{\ast}\!(\Theta ,\Phi)
\ Y_{\ell m}^{\ast}\!(\Theta ,\Phi) \cdot  \nonumber  \\
                                            & & \cdot \int _{0}^{\pi} 
\int _{0}^{2\pi} Y_{11}\!(\theta ,\phi)\ Y_{\ell m}\!(\theta ,\phi)\ 
Y_{\ell ' m'}^{\ast}\!(\theta ,\phi)\ \sin\!\theta \ 
{\mathrm{d}}\theta \ {\mathrm{d}}\phi .   
\label{eq:B8}      
   \end{eqnarray}   
In each of (\ref{eq:B6})-(\ref{eq:B8}), we decompose the spherical 
harmonics into associated Legendre functions 
(formula (3.53) of Jackson \shortcite{Jack}), 
perform the integration over $\theta$ and $\phi$, utilising 
formulae (8.733.2)-(8.733.4) of Gradshteyn \& Ryzhik \shortcite{GraRy} and the 
orthogonality condition of the associated Legendre functions 
(formula (3.52) of Jackson \shortcite{Jack}), and find these expressions 
to give
   \begin{eqnarray}
       \stackrel{1}{{\mathrm{g}}}_{\ell ' m'}&=&\frac{4\pi}{(2\pi)^{3/2}}\ 
{\mathrm{i}} ^{\ell ' +2}\ \int _{-\infty}^{+\infty} 
{\mathrm{d}}^{3}{\mathbf{n}}\ 
B\!_{\mathbf{n}}\ n(1- \xi)\ \sqrt{\frac{2\ell '+1}{4\pi}\ 
\frac{(\ell '-m')!}{(\ell '+m')!}}\ \frac{1}{2\ell '+1} \cdot \nonumber \\ 
                                           & &\cdot \left[ j_{\ell '+1}\!
(n\xi)\ \frac{\sqrt{1-{\mathcal{B}}^{2}}}{2}\ P_{\ell '+1}^{m'+1}\!
({\mathcal{B}}) + j_{\ell '-1}\!(n\xi)\ \frac{\sqrt{1-{\mathcal{B}}^{2}}}{2}
\ P_{\ell '-1}^{m'+1}\!({\mathcal{B}}) \right]\ 
{\mathrm{e}}^{-{\mathrm{i}} m'\Phi}     
\label{eq:B9}
   \end{eqnarray}
   \begin{eqnarray}
       \stackrel{2}{{\mathrm{g}}}_{\ell ' m'}&=&\frac{4\pi}{(2\pi)^{3/2}}\ 
{\mathrm{i}} ^{\ell ' +2} \int _{-\infty}^{+\infty} 
{\mathrm{d}}^{3}{\mathbf{n}}\ 
B\!_{\mathbf{n}}\ n(1- \xi)\ \sqrt{\frac{2\ell '+1}{4\pi}\ \frac{(\ell '-m')!}
{(\ell '+m')!}}\ \frac{1}{2\ell '+1} \cdot \nonumber \\ 
                                           & &\cdot \left[ j_{\ell '+1}\!(n\xi)
\ [-(\ell ' -m'+1)]\ {\mathcal{B}}\ P_{\ell '+1}^{m'}\!({\mathcal{B}}) + 
j_{\ell '-1}\!(n\xi)\ (\ell '+m')\ {\mathcal{B}}\ P_{\ell '-1}^{m'}\!
({\mathcal{B}}) \right]\ {\mathrm{e}}^{-{\mathrm{i}} m'\Phi}     \label{eq:B10}
   \end{eqnarray}
   \begin{eqnarray}
       \stackrel{3}{{\mathrm{g}}}_{\ell ' m'}&=&\frac{4\pi}{(2\pi)^{3/2}}\ 
{\mathrm{i}} ^{\ell ' +2}\ \int _{-\infty}^{+\infty} 
{\mathrm{d}}^{3}{\mathbf{n}}\ 
B\!_{\mathbf{n}}\ n(1- \xi)\ \sqrt{\frac{2\ell '+1}{4\pi}\ \frac{(\ell '-m')!}
{(\ell '+m')!}}\ \frac{1}{2\ell '+1} \cdot \nonumber \\ 
                                           & &\cdot \left[ j_{\ell '+1}\!(n\xi)
\ \frac{-(\ell ' -m'+1)(\ell ' -m'+2)}{2}\ \sqrt{1-{\mathcal{B}}^{2}}\ 
P_{\ell '+1}^{m'-1}\!({\mathcal{B}}) + \right. \nonumber \\
                                           & &+ \left. j_{\ell '-1}\!(n\xi)\ 
\frac{-(\ell '+m'-1)(\ell '+m')}{2}\ \sqrt{1-{\mathcal{B}}^{2}}\ 
P_{\ell '-1}^{m'-1}\!({\mathcal{B}}) \right]\ 
{\mathrm{e}}^{-{\mathrm{i}} m'\Phi}   ,  
\label{eq:B11}
   \end{eqnarray}
where we have used the shorthand notations  
\[
{\mathcal{A}} \equiv \cos\!\theta,\   {\mathcal{B}} \equiv \cos\!\Theta.      
\]
Substituting (\ref{eq:B9})-(\ref{eq:B11}) in (\ref{eq:B5}), and using 
formulae (8.733.2)-(8.733.4) of Gradshteyn \& Ryzhik \shortcite{GraRy} to 
show that 
\[
\frac{1}{2}\sqrt{1-{\mathcal{B}} ^{2}}\ P_{\ell +1}^{m+1}\!({\mathcal{B}}) -
(\ell -m+1)\ {\mathcal{B}}\ P_{\ell +1}^{m}\!({\mathcal{B}}) -
\frac{1}{2}(\ell -m+1)(\ell -m+2)\ \sqrt{1-{\mathcal{B}} ^{2}}\ 
P_{\ell +1}^{m-1}\!({\mathcal{B}})=-(\ell +1)\ P_{\ell}^{m}\!({\mathcal{B}})
\]
and
\[
\frac{1}{2}\sqrt{1-{\mathcal{B}} ^{2}}\ P_{\ell -1}^{m+1}\!({\mathcal{B}}) +
(\ell +m)\ {\mathcal{B}}\ P_{\ell -1}^{m}\!({\mathcal{B}}) -
\frac{1}{2}(\ell +m-1)(\ell +m)\ \sqrt{1-{\mathcal{B}} ^{2}}\ 
P_{\ell -1}^{m-1}\!({\mathcal{B}})= \ell\ P_{\ell}^{m}\!({\mathcal{B}}),
\]
we finally arrive at (\ref{eq:B1}).

\end{document}